\documentclass[twoside]{article} 
\usepackage[accepted]{aistats2025}
\usepackage[round]{natbib}

\usepackage[utf8]{inputenc} 
\usepackage[T1]{fontenc}    
\usepackage[hidelinks]{hyperref}       
\usepackage{url}            
\usepackage{booktabs}       
\usepackage{amsfonts}       
\usepackage{nicefrac}       
\usepackage{microtype}      
\usepackage[dvipsnames]{xcolor}         
\hypersetup{
  colorlinks   = true, 
  urlcolor     = blue, 
  linkcolor    = blue, 
  citecolor   = red 
}
\usepackage{caption}
\usepackage{changepage}
\usepackage{amssymb}
\usepackage{amsmath}
\usepackage{amsthm}
\newtheorem{theorem}{Theorem}
\newtheorem{lemma}{Lemma}

\usepackage{graphicx}
\usepackage{float}
\usepackage{algorithm}
\usepackage{algpseudocode}
\usepackage{graphicx} 
\usepackage{color,soul}

\begin{document}

%

%

\twocolumn[

\aistatstitle{Parameter Estimation of State Space Models Using Particle Importance
Sampling}

\aistatsauthor{ Yuxiong Gao \And Wentao Li \And Rong Chen}

\aistatsaddress{ University of Manchester \\ Manchester, UK \And University of Manchester \\ Manchester, UK \And Rutgers University \\ New Jersey, USA} ]

\begin{abstract}
 State-space models have been used in many applications, including econometrics, engineering, medical research, etc. The maximum likelihood estimation (MLE) of the static parameter of general state-space models is not straightforward because the likelihood function is intractable. It is popular to use the sequential Monte Carlo(SMC) method to perform gradient ascent optimisation in either offline or online fashion. One problem with existing online SMC methods for MLE is that the score estimators are inconsistent, i.e. the bias does not vanish with increasing particle size. In this paper, two SMC algorithms are proposed based on an importance sampling weight function to use each set of generated particles more efficiently. The first one is an offline algorithm that locally approximates the likelihood function using importance sampling, where the locality is adapted by the effective sample size (ESS). The second one is a semi-online algorithm that has a computational cost linear in the particle size and uses score estimators that are consistent. We study its consistency and asymptotic normality. Their computational superiority is illustrated in numerical studies for long time series.    
\end{abstract}

\section{INTRODUCTION}
State-space models are a class of stochastic models in which the observations $\{y_{t}\}_{t=1}^T$ are based on unobservable latent state variables $\{x_{t}\}_{t=1}^T$ through some probabilistic dependence. This class of models has been extensively studied due to its wide applications in science and engineering. In this paper, we mainly consider the state-space models in discrete time and have Markov property:
\begin{equation}   
 \label{ssm}
x_{t}|x_{0:t-1} \sim f_{\theta}(\cdot|x_{t-1}),\;
y_{t}|x_{0:t},y_{0:t-1} \sim g_{\theta}(\cdot|x_{t}),
\end{equation}
where $t\in \mathbb{Z^{+}}$, $x_{0:t}$ refers to $x_0,\dots, x_t$, $\theta$ is the parameter vector, $f_{\theta}(x_t|x_{t-1})$ and $g_{\theta}(y_t|x_{t})$ are probability densities. The question of interest here is the true parameter $\theta^\star$. Since $\{x_{t}\}_{t=1}^T$ are not observable, the likelihood function $p_{\theta}(y_{0:T})$ is intractable in the sense that a direct attempt to calculate the likelihood function may involve high-dimensional integration over $\{x_{t}\}_{t=1}^T$, except for the linear Gaussian state-space model  \citep{kalman1960new}. Sequential Monte Carlo (SMC) is a popular class of methods that approximate the likelihood by generating Monte Carlo samples, so-called `particles', for the latent variables and assigning them with weights corresponding to the sequential importance sampling with resampling \citep{andrieu2005arnaud}. This work focuses on the MLE of the true parameter value $\theta^\star$.

Given a parameter value $\theta$, the SMC method provides particles targeting the conditional distribution $p_{\theta}(x_{0:T}| y_{0:T})$ \citep{kantas2015particle} and the estimated likelihood \citep{douc2008limit}. It is well-known that the estimated likelihood function is not continuous due to the independent Monte Carlo noise on each SMC run and the discreteness of the resampling step \cite[]{doucet2023differentiable}. 
To apply gradient-based optimizations, the score function is often estimated using Fisher's identity when the gradients of the state and observation densities can be evaluated easily, as given below,
{\small
\begin{equation}
\nabla_{\theta} l(\theta)=\int \nabla_{\theta} \log p_\theta(x_{0: T}, y_{0: T}) p_\theta\left(x_{0: T} | y_{0: T}\right)\, d x_{0: T}, \label{Fisher_ident}
\end{equation}
}
where the log-likelihood function  $l(\theta)\triangleq \log p_{\theta}(y_{0:T})$. 
Given particles from $p_\theta\left(x_{0:T}|y_{0: T}\right)$, \eqref{Fisher_ident} can be estimated consistently and the steepest gradient ascent (SGA) can be used. Alternatively, \eqref{Fisher_ident} can be estimated using a recursive expression without needing to store the full paths of particles in the cost of $O(N^2)$ computational complexity \citep{poyiadjis2011particle}. However, both methods require a new set of particles to be generated for different $\theta$, and the computational cost is sensitive to the optimization scheme. For example, particles may be repeatedly generated for similar parameter values if the step size is small, or for similar likelihood values in regions where parameter values are close to unidentifiable. 

Sometimes it is more efficient to update $\theta$ when processing the data on the fly, e.g. when the early part of the observed series is informative about the parameter. Based on the decomposition $p_\theta(y_{0: T})=\prod_{t=1}^Tp_\theta(y_t| y_{0:t-1})p_\theta(y_0)$ and the ergodicity assumption on $y_{0:T}$, by iterating the following update from $t=0$ to $t=T-1$,
\begin{equation}
\theta_{t+1}=\theta_t+\gamma_{t} \nabla_{\theta} \log p_\theta(y_t| y_{0: t-1})|_{\theta=\theta_t}, \label{online_update}
\end{equation}
it is expected that $\theta_T$ will have the same asymptotic behaviour as the MLE. The challenge is that consistent estimation of the conditional score function $\nabla_{\theta} \log p_\theta(y_t| y_{0: t-1})$ using SMC requires particles simulated with the parameter value $\theta_t$  different at each time $t$, and regenerating particles at every $t$ is contrary to the purpose of using \eqref{online_update} for the online update. Existing online SMC implementations of \eqref{online_update} bypass this problem by propagating the particles under the `time-varying' parameter $\theta_{0:n}$   \citep{poyiadjis2006gradient,poyiadjis2011particle,nemeth2016particle}. For finite state-space hidden Markov models, it was shown that such an algorithm converges toward $\theta^{\star}$ under regularity conditions   \citep{legland1997recursive}. For general state-space models, however, there is no theoretical support and it contains a non-vanishing bias as $N$ increases. It is also sensitive to the initialisation of $\theta$ which needs to be close to the true parameter $\theta^{\star}$. This is illustrated numerically in Section 4.1. 

This paper proposes to recycle generated particles for performing multiple updates in SGA using \eqref{Fisher_ident} and a novel semi-online SMC algorithm that performs \eqref{online_update} under varying parameter values with consistent estimators, exploiting the fact that for $x_{0:t}\sim p_{\theta_0}(x_{0:t}| y_{0:t})$, $x_{0:t}$ weighted with the importance sampling weight, defined as 
\begin{equation}
a_{\theta_{0}}(\theta,x_{0:t}) \triangleq p_{\theta}(x_{0:t},y_{0:t})/p_{\theta_{0}}(x_{0:t},y_{0:t}),	
\end{equation}
 follows the density $p_{\theta}(x_{0:t}| y_{0:t})$. Numerical experiments show that the new methods significantly improve the standard SGA by up to one order of magnitude, and are competitive to existing state-of-the-art algorithms. There are three main contributions. First, for the offline implementation, the score function in a neighbourhood around $\theta_0$ is approximated using particles weighted with $a_{\theta_{0}}(\theta,x_{0:t})$, and the neighbourhood is determined by thresholding ESS of $a_{\theta_{0}}(\theta,x_{0:t})$ to avoid the poor approximation when $\theta$ is far way from $\theta_{0}$. Second, a combination of the online gradient ascent and a SMC algorithm that renews part of the particle path is proposed, which provides a consistent estimator of the conditional score function and has a computational complexity of $O(N)$. Specifically, at the $k$th iteration, the weights of particles targeting $p_{\theta_{k-1}}(x_{0:k}| y_{0:k})$ are multiplied with $a_{\theta_{k-1}}(\theta_k,x_{0:k})$, so that the weighted particles are `retargeted' to the new parameter value. A renewing step is introduced in which, if the quality of the particles is below a certain threshold, an SMC algorithm is run to regenerate particles targeting $p_{\theta_{k}}(x_{0:k}| y_{0:k})$. Third, the consistency and asymptotic normality of the conditional score estimator as $N$ goes to infinity is stated, which provide justification for the form of the thresholding ESS in the renewing step.

\paragraph{Notations}
Capital letters such as $X,Y$ represent random variables and lowercase $x,y$ their corresponding values. The latent variable is denoted by $X\in \mathcal{X}$ and the observation by $Y\in \mathcal{Y}$. The sequence of latent variables $x_{t},\dots,x_{t'}$ is denoted by $x_{t:t'}$, and similarly the observations $y_{t},\dots,y_{t'}$ by $y_{t:t'}$, where $t'\geq t$ are non-negative integers. The static parameter $\theta \in \mathbb{R}^p$ in \eqref{ssm} is of interest. The density of the initial latent variable $X_{0}$ at value $x_{0}$ is denoted by $f_{\theta}(x_{0})$. For any function $h$ of $\theta$ we denote its gradient with respect to its argument $\theta$ as $\nabla h(\theta)$, the $i^{th}$ coordinate of the gradient as $\nabla _{i}h(\theta)$, and the value of the gradient at $\theta_{0}$ as $\nabla h(\theta)|_{\theta=\theta_{0}}$ or $\nabla h(\theta_{0})$. 
\subsection{Maximum Likelihood Estimation of State-space Models Using SMC}

At each time point $t$, for a given $\theta$ and $t_0<t$, the conditional likelihood $p_{\theta}(y_{(t_0+1):t}\mid y_{0:t_0})$ can be written by the Markovian structure as follows, 
{\small
\begin{equation}
\frac{p_\theta(y_{0:t})}{p_{\theta}(y_{0:t_0})}=\int p_{\theta}(x_{(t_0+1):t}, y_{(t_0+1): t} | x_{t_0}) p_{\theta}(x_{t_0} | y_{0:t_0})d x_{t_0:t},   \label{ss_likelihood} 	
\end{equation}
}
where $p_{\theta}(x_{(t_0+1):t}, y_{(t_0+1): t}|x_{t_0})$ can be factorised as $\prod_{j=t_0}^{t-1} f_{\theta}(x_{j+1}|x_j)g_{\theta}(y_{j+1}|x_{j+1})$. An online gradient ascent to approximately maximise $p_{\theta}(y_{0:T})$ using the SMC method on \eqref{online_update} is given in Algorithm \ref{onlineMLE}. If the step size $a_t\equiv0$, the parameter value is kept fixed and Algorithm \ref{onlineMLE} reduces to the vanilla SMC algorithm which can be used for offline MLE with \eqref{Fisher_ident}.

\setlength{\textfloatsep}{13pt}

\begin{algorithm}[h]
  \caption{Online gradient ascent / vanilla SMC}
\label{onlineMLE}
  \begin{algorithmic}
   \State
Assume that the initial value $\theta_0$, the threshold $r_2\in(0,1)$ and the step sizes $\{\gamma_{t}\}_{n=0}^{T}$ are given. At $t=0$, the proposal distribution is $q_{0}(x_{0})$. Let $\widetilde{w}_{-1}^{(i)}=1$, $i=1,\cdots,N$.\\
\noindent \textbf{For} $t = 0$ to $T-1$:
\begin{adjustwidth}{1cm}{0cm}
\noindent\textbf{Propagation \& Weighting:}
For $i=1,\cdots,N$, sample $x_{t}^{(i)}\sim q_{t}(x_{t}|\widetilde{x}_{0:t-1}^{(i)})$,\\ calculate $u_{t}^{(i)}=\frac{p_{\theta_t}(x_{t}^{(i)},y_{t}|\widetilde{x}_{0:t-1}^{(i)})}{q_{t}(x_{t}^{(i)}|\widetilde{x}_{0:t-1}^{(i)})}$ and let $w_{t}^{(i)}=u_{t}^{(i)}\widetilde{w}_{t-1}^{(i)}$, $x_{0:t}^{(i)}=(\widetilde{x}_{0:t-1}^{(i)},x_{t}^{(i)})$.\\ 
\noindent\textbf{Conditional GA:}
Set $\theta_{t+1}=\theta_{t}+\gamma_{t} \widehat{\nabla}\mathrm{log} p_{\theta}(y_t|y_{0:t-1})|_{\theta=\theta_{t}}$.

\noindent\textbf{Resampling:}
If ${\rm ESS}(w_t^{1:N})/N\leq r_{2}$, resample $\{x_{0:t}^{(i)}\}_{i=1}^{N}$ with weights $\{w_t^{(i)}\}_{i=1}^{N}$ to get $\{\widetilde{x}_{0:t}^{(i)}\}_{i=1}^{N}$ and set all $\widetilde{w}_t^{(i)}=1$. Otherwise set $\{(\widetilde{x}_{0:t}^{(i)},\widetilde{w}_{t}^{(i)})\}_{i=1}^N=\{(x_{0:t}^{(i)},w_{t}^{(i)})\}_{i=1}^N$.
\end{adjustwidth}
\textbf{End For}
\end{algorithmic}
\end{algorithm}
The ESS above is defined as $ESS(w_{1:N})\triangleq 1/\sum_{i=1}^{N}\overline{w}_i^{2}$ for a set of importance weights $w_{1:N}$, where $\overline{w}_i$ is the normalised weights defined as $w_{i}/\sum_{j=1}^{N}w_{j}$ \citep{martino2017effective}. Algorithm \ref{onlineMLE} is taken by the literature which propose estimators of $\nabla\mathrm{log}p_{\theta}(y_{t+1}|y_{0:t})$ for the online updates   \citep{chopin2020introduction,nemeth2016particle,poyiadjis2011particle}. When $\theta_t\equiv\theta$, at each time $t$, the weighted particles $(x_{0:t}^{(i)},w_{t}^{(i)})$ and $(\widetilde{x}_{0:t}^{(i)},\widetilde{w}_{t}^{(i)})$ approximately follow the density $p_{\theta}(x_{0:t}|y_{0:t})$, $i=1,\cdots,N$. Therefore \eqref{ss_likelihood} can also be estimated asymptotically unbiasedly using the particles, and  $l(\theta)$ can be estimated by $\widetilde{l}(\theta)\triangleq \Sigma_{j=1}^{k+1} \mathrm{log}(N^{-1}\Sigma_{i=1}^{N}w_{t_{j}}^{(i)})$ where $t_1,\cdots,t_{k+1}$ are the times when resampling is performed. 

However, when $\theta_t$ is updated online, the particles instead follow the density $p_{\theta_{0:t}}(x_{0:t}|y_{0:t})$ where the parameter value of the state and observation densities varies over $t$. For example, at time $t=1$, the particle $x_{1:2}^{(i)}$ weighted with $w_2^{(i)}$ targets the density proportional to $p_{\theta_1}(x_2,y_2|x_1)p_{\theta_0}(x_1|y_1)$, which means the marginal density of weighted $x_2^{(i)}$ is biased towards the filtering density of $x_2$ at the parameter value $\theta_1$. Thus, the conditional score cannot be estimated consistently and the estimated MLE contains a non-vanishing bias. Algorithm \ref{onlineMLE} seems to perform well empirically in the literature, but requires the initial parameter $\theta_0$ to be close to the true value which is usually found using the a short segment of the data. 


On the other hand, the development of Bayesian inference methods for the static parameter is very active, including both online and offline methods \citep{carvalho2010particle,andrieu2010particle,chopin2013smc2,rosato2022efficient}. One method of updating the posterior distribution of $\theta$ online is to approximate the likelihood at the proposed $\theta_t$ using that at $\theta_{t-1}$ by controlling $\|\theta_t-\theta_{t-1}\|$ to be small \cite[]{crisan2018nested,crisan2017uniform,perez2018probabilistic}. It is not in our purpose to promote MLE over Bayesian inference of the static parameter, but to provide an alternative to existing offline and online SMC MLE estimators for users who prefer an alternative to choosing a prior distribution for $\theta$ and tuning and assessing the mixing Markov chain sampler when performing Bayesian inference.

\section{NEW ALGORITHMS}
\subsection{Adaptive Gradient Ascent Using ESS}

Following the importance sampling form below, 
\begin{equation}
\frac{p_{\theta}(y_{0:T})}{p_{\theta_0}(y_{0:T})}= \int a_{\theta_{0}}(\theta,x_{0:T})p_{\theta_{0}}(x_{0:T}|y_{0:T})\,dx_{0:T}, \label{lre}	
\end{equation}
the Fisher's identity \eqref{Fisher_ident} can be extended to give the score estimator 
$$
\widehat{\nabla}l_{\theta_{0}}(\theta)\triangleq\frac{\sum_{i=1}^{N}\nabla\log p_\theta(x_{0: T}^{(i)}, y_{0: T})a_{\theta_{0}}(\theta,x_{0:T}^{(i)})w_{T}^{(i)}}{\sum_{i=1}^{N}a_{\theta_{0}}(\theta,x_{0:T}^{(i)})w_{T}^{(i)}}.
$$ 

The approximation is accurate when $\theta$ is close to $\theta_{0}$, but unreliable when $\theta$ is far away, because particles approximating the distribution $p_{\theta_{0}}$ may not cover the high-density region of $p_{\theta}$. It is natural to use $\widehat{\nabla}l_{\theta_{0}}(\theta)$ in the neighbourhood of $\theta_0$ adaptively determined by the quality of the importance weights. Based on this we introduce an offline algorithm in Algorithm \ref{offline_algo} that generalises the SGA .

\begin{algorithm}[h]
  \caption{Adaptive gradient ascent with particle importance sampling (adaptGA-PIS)}
\label{offline_algo}
  \begin{algorithmic}
   \State
Assume that the initial value $\theta_0$, a threshold $r\in(0,1)$ and the step sizes $\{\gamma_{n}\}_{n=0}^{I}$ are given.\\
\noindent \textbf{For} $n = 0$ to the maximum iterations $I$:
\begin{enumerate}
	\item Run an SMC algorithm from $t=0$ to $T$ with parameter $\theta_n$ to obtain the weighted particles $\{(x_{0:T}^{(i)},w_T^{(i)})\}_{i=1}^N$. Let $\theta_{n,1}=\theta_n$ and $k=1$.
	\item \textbf{While} $\mbox{ESS}(\{a_{\theta_n}(\theta_{n,k},x_{0:T}^{(i)})\}_{i=1}^N)>rN$ and $\theta_{n,k}$ does not converge:
Let $\theta_{n,k+1}=\theta_{n,k}+\gamma_{n}\widehat{\nabla}l_{\theta_{n}}(\theta)|_{\theta=\theta_{n,k}}$, where $\widehat{\nabla}l_{\theta_{n}}(\theta)$ estimates $\nabla l(\theta)$ using \eqref{lre}, and $k=k+1$. 
	\item Let $\theta_{n+1}=\theta_{n,k}$. Stop if $\theta_n$ converges, otherwise go to step 1.
\end{enumerate}	
\noindent\textbf{End For}
\end{algorithmic}
\end{algorithm}

One commonly used SMC algorithm is given in the appendix. The value of ESS is equal to $1$ when the weights are all equal and close to $0$ when a few weights dominate the others, which, in our context, corresponds to $\theta=\theta_0$ and $\theta$ is far away from $\theta_0$. We provide more details on the choice of step sizes $\gamma_{n}$ and the tuning for $r$ in Appendix D.  

If in step 2 ESS is removed from the stopping rules, the algorithm reduces to the Monte Carlo maximum likelihood (MCML) \citep{geyer1992constrained} or the SGA if step 2 ends at $k=2$. Compared to MCML, the unreliable score estimates are avoided. Compared to the SGA, the additional computational cost of Algorithm \ref{offline_algo} is negligible when the cost of evaluating $\{a_{\theta_n}(\theta,x_{0:T}^{(i)})\}_{i=1}^N$ is negligible compared to that of generating particles. In this case, Algorithm \ref{offline_algo} is computationally more efficient since one set of particles is used for multiple gradient ascent steps, particularly when small step sizes are used, e.g. in the vicinity of MLE.

When $T$ is large, intuitively it is sufficient that $\theta_{n,k}-\theta_n=O(T^{-1/2})$ for the ESS to be away from $0$ by noting the following: in the expression of $\widehat{\nabla}l_{\theta_{n}}(\theta)$, by the central limit theorem over the weighted particles, the normalised importance weight can be expanded as follows,
\vspace{-2pt}
\begin{align}
 & \frac{a_{\theta_{n}}(\theta,x_{0:T})}{\sum_{i=1}^{N}a_{\theta_{n}}(\theta,x_{0:T}^{(i)})w_{T}^{(i)}} =\frac{p_{\theta}(x_{0:T}|y_{0:T})}{p_{\theta_{n}}(x_{0:T}|y_{0:T})}\{1+ O_p\left(1/\sqrt{N}\right)\} \nonumber\\
 & =\exp\left[\left\{ \nabla_{\theta}\log p_{\dot{\theta}}(x_{0:T},y_{0:T})-\nabla_{\theta}\log p_{\dot{\theta}}(y_{0:T})\right\} (\theta-\theta_{n})\right] \nonumber \\
 &\cdot\{1+ O_p\left(1/\sqrt{N}\right)\},\label{taylorapprox}
\end{align}
\noindent which is in the order of $\exp\{O_{p}(\sqrt{T}(\theta-\theta_{n}))\}$, since the two score functions have the form of summation and are in order of $\sqrt{T}$ under standard regularity conditions for the state-space model. The second equality of \eqref{taylorapprox} holds by the Taylor expansion and $\dot{\theta}$ is an intermediate value between $\theta_{n}$ and $\theta$. Since the gradient and Hessian of the log-likelihood function typically has the order $O(\sqrt{T})$ and $O(T)$ respectively \cite[]{durbin2012time,hamilton2020time}, one set of particles can be recycled for multiple times within the range of a Newton-Raphson iteration. This is illustrated in the numerical experiments with $T=10000$. 

\begin{figure}[h]
  \centering
\includegraphics[width=0.5\textwidth]{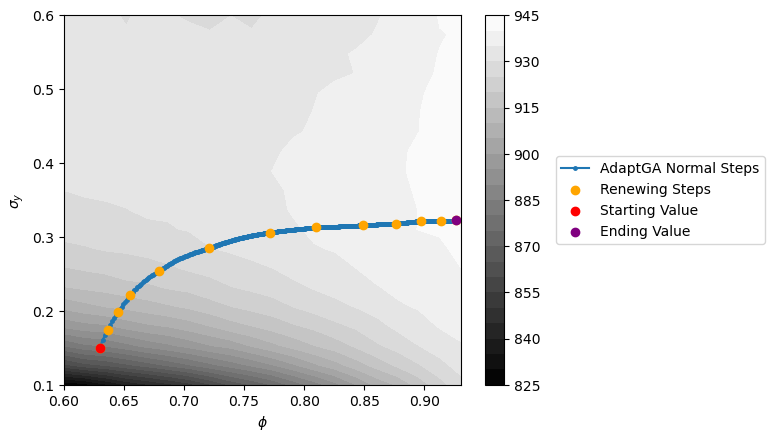}
\includegraphics[width=0.5\textwidth]{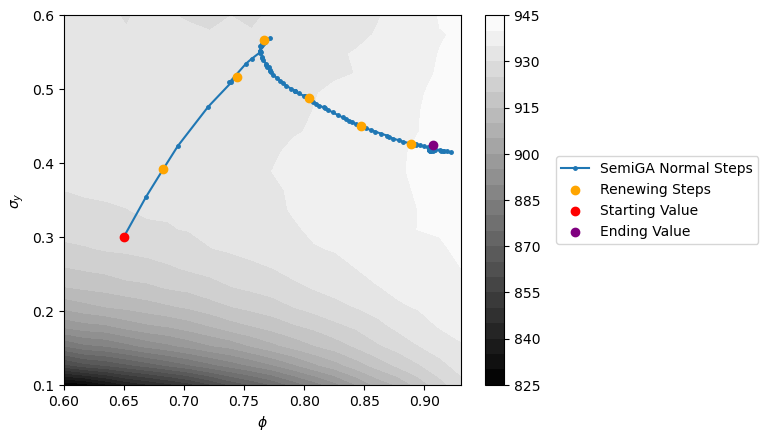}
  \caption{\small{Trajectory plots for one replication of the algorithms run on the model \eqref{sv}. Here $T=100$ and $\sigma_{x}=0.4$ is known, so the parameter of interest is $\theta \triangleq (\phi,\sigma_{y})$. The plot on the top is the trajectory for AdaptGA-PIS and the plot on the bottom is the trajectory for SemiGA-PIS. The initial values are $\theta_{0}=(0.63,0.15)$ and $\theta_{0}=(0.65,0.3)$ respectively. The MLE estimated by a sufficient number of Fisher SGA is $\theta^{\star}=(0.91,0.39)$. Full settings can be found in the code.} }
\label{sv_surface}
\end{figure}

It is challenging to use the particle-based SGA if some parameters are hardly identifiable, because $l(\theta)$ is non-concave and flat in some region of the parameter space. The SGA may repeatedly generate particles at parameter values where the likelihood changes little, so its computational cost is sensitive to the initialisation and the choice of step size. In contrast, Algorithm \ref{offline_algo} renews the particles less in such regions as the joint density $p_{\theta}(x_{0:T},y_{0:T})$ is not sensitive to the change of $\theta$, hence is more robust to the choice of step size by adapting to the curvatures. See Figure \ref{sv_surface} for an illustration. 



\subsection{Semi-online Gradient Ascent Controlled by ESS}

In Algorithm \ref{offline_algo}, each gradient ascent step is conditional on the entire sequence of data. If the data is long or the early part of the data is informative about the parameter, it may be cheaper to update the parameter value using \eqref{online_update}. At each iteration which is also time $t$, to obtain particles following the filtering density at the latest parameter value, we propose to adjust the particle weight with $a_{\theta_{t}}(\theta_{t+1}, x_{0:t})$, which gives Algorithm \hyperlink{1}{3}. 

\begin{algorithm}
  \caption{Semi-online gradient ascent with particle importance sampling (semiGA-PIS)}
\label{onlineAIS}
  \begin{algorithmic}
   \State
Assume that the initial value $\theta_0$, thresholds $r_1,r_2\in(0,1)$ and the step sizes $\{\gamma_{t}\}_{t=0}^{T}$ are given. At $t=0$, the proposal distribution is $q_{0}(x_{0})$. Let $\widetilde{w}_{-1}^{(i)}=1$, $i=1,\cdots,N$.\\
\noindent \textbf{For} $t = 0$ to $T-1$:
\begin{adjustwidth}{0.5cm}{0cm}
\noindent\textbf{Propagation \& Weighting:} Same as that in Algorithm \ref{onlineMLE}.\\ 
\noindent\textbf{Conditional GA:}\\
Set $\theta_{t+1}=\theta_{t}+\gamma_{t} \widehat{\nabla}\mathrm{log}p_{\theta}(y_t|y_{0:t-1})|_{\theta=\theta{t}}$. \\

\noindent\textbf{Retarget:}
Multiply $w_{t}^{(i)}$ by $a_{\theta_{t}}(\theta_{t+1},x_{0:t}^{(i)})$ to obtain $\widetilde{w}_{t}^{(i)}$, $i=1,\cdots,N$.\\
\noindent\textbf{Renewing:}\\
If $\sum_{k=t-K+1}^t\mbox{ESS}(a_{\theta_{k}}(\theta_{k+1},x_{0:t}^{1:N}))/N \leq r_{1}$, run an SMC algorithm to obtain a new set of particles targeting $p_{\theta_{t+1}}(x_{0:t}|y_{0:t})$, denoted by $\{\widetilde{x}_{0:t}^{(i)}\}_{i=1}^{N}$. Set all $\widetilde{w}_{t}^{(i)}=1$.\\
\noindent\textbf{Resampling:}
If $\mbox{ESS}(\widetilde{w}_{t}^{1:N})/N\leq r_{2}$, resample $\{x_{0:t}^{(i)}\}_{i=1}^{N}$ with weights $\{\widetilde{w}_{t}^{(i)}\}_{i=1}^{N}$ to get $\{\widetilde{x}_{0:t}^{(i)}\}_{i=1}^{N}$ and set all $\widetilde{w}_{t}^{(i)}=1$. Otherwise set $\{\widetilde{x}_{0:t}^{(i)}\}_{i=1}^N=\{x_{0:t}^{(i)}\}_{i=1}^N$.

\end{adjustwidth}
\textbf{End For}
\end{algorithmic}
\end{algorithm}
The conditional GA step is the same as that in Algorithm \ref{onlineMLE} and the same conditional score estimator therein can be used. In our experiment, both algorithms use the difference $\widehat{\nabla}\log p_{\theta}(y_{0:t})-\widehat{\nabla}\log p_{\theta}(y_{0:(t-1)})$ where each partial score is estimated using \eqref{Fisher_ident}. Compared to Algorithm \ref{onlineMLE}, it adds the retarget step which gives $\{x_{0:t}^{(i)},\widetilde{w}_{t}^{(i)}\}_{i=1}^{N}$ approximately following $p_{\theta_{t+1}}(x_{0:t}|y_{0:t})$. Thus, at each iteration, the propagation and weighting step is conditional on particles from the filtering density having the correct parameter value. Unlike Algorithm \ref{onlineMLE}, the new algorithm gives a consistent estimator of $\nabla\mathrm{log}p_{\theta}(y_{t}|y_{0:t-1})|_{\theta=\theta{t}}$ as $N\rightarrow\infty$, whereas that in Algorithm \ref{onlineMLE} is inconsistent. For the tuning of $r_{1}$ see Appendix D for more details. 

At iteration $t$, if $\theta_{t+1}$ is very different from some earlier parameter values $\theta_k$, $k<t+1$, the particles may suffer from particle degeneracy. Because $x_k^{(i)}$ generated with $\theta_k$ when conditioned on $y_{0:k}$ may be in tail areas of $p_{\theta_{t+1}}(x_k|y_{0:t})$, when the information of later observations are included. The renewing step is introduced to regenerate the entire particle trajectory up to the current iteration if the particles degenerate severely due to the changes in parameter values. This degeneracy is different from that caused by the weighting step and is monitored by the quality of $a_{\theta_{t}}(\theta_{t+1},x_{0:t})$. To see this, note that when $\theta_k$ is fixed, there is still particle degeneracy but no regeneration is performed. This can also be seen in \eqref{eq:asy_var_term}. The ESS of $a_{\theta_{t}}(\theta_{t+1},x_{0:t}^{1:N})$ measures the variation of changes of $p_{\theta_t}(x_{0:t}^{(i)},y_{0:t})$ when $\theta_t$ is changed to $\theta_{t+1}$ and controls the resulting additional variance that is added to the estimator $\widehat{\nabla}\mathrm{log}p_{\theta}(y_{t+1}|y_{0:t})|_{\theta=\theta_{t+1}}$. When particles are in tail areas of $p_{\theta_t}(x_{0:t}|y_{0:t})$, the change is likely to vary greatly and trigger the renewal of  particles. Between two renewals, early part of the particles usually degenerate under repeated resampling. Denote $t_0$ as the largest index where $x_{0:t_0}^{(i)}$ have an identical value denoted by $x_{0:t_0}^{(0)}$. If $t_0>0$, the ESS of $a_{\theta_{t}}(\theta_{t+1},x_{0:t}^{1:N})$ measures the variation of changes of $p_{\theta_t}(x_{(t_0+1):t}^{(i)},y_{(t_0+1):t}|x_{0:t_0}^{(0)},y_{0:t_0})$ instead. When the particles are in tail areas, the skewness can still be detected with sufficient diversity on the dimensions    $x_{(t_0+1):t}$ despite the coalescence on the dimensions $x_{0:t_0}$.  In the renewing step, the average of the most recent several ESS values is used for stability purpose. 

Both Algorithm \ref{onlineMLE} and the online part of Algorithm \ref{onlineAIS} cost $O(TN)$ for length-$T$ data. The new algorithm is less sensitive to the initial value $\theta_0$ than Algorithm \ref{onlineMLE} and gives the consistent estimator by using the offline renewal with the additional cost $O(\sum_j^dt_jN)$ if it occurs at time $t_1,\cdots,t_d$, hence it is identified as `semi-online'. However, it does not mean Algorithm \ref{onlineAIS} has an overall cost of $O(T^{2}N)$, since the renewal steps are not necessarily a recurrent or regular process proportional to iteration $T$. The frequency depends on $r_1$ and the sensitivity of the joint density to the change of the parameter value. The renewal occurs relatively frequently in the early iterations where the gradients are of a considerable scale and only on an occasional basis when $\theta$ is in close proximity to the MLE where the gradient is relatively small. Algorithm \ref{onlineMLE} has the following benefits. First, similarly to Algorithm \ref{offline_algo}, the frequency of regenerating the particle trajectories is adaptive to the likelihood surface, but in an online manner, by being less frequent in the region where the joint density is flat. Second, Algorithm \ref{onlineAIS} does not need a pilot phase for initialisation which Algorithm \ref{onlineMLE} needs. Third, if $p_{\theta}(x_{0:t},y_{0:t})$ depends on the latent states through a fixed $d$-dimensional summary statistics as that of models considered in \cite{fearnhead2002markov}, only a $d\times N$ matrix needs to be stored and the $O(tN)$ cost at iteration t can be avoided, preventing the total cost from being $O(T^2N)$. These are illustrated in the numerical studies with large
T values.



\section{ASYMPTOTIC PROPERTY}
For Algorithm \ref{offline_algo}, since the weighted particles are from some existing SMC algorithm, $\widehat{\nabla}l_{\theta_{n}}(\theta)$ satisfies the corresponding central limit theorem  \citep{chopin2004central}, and has a finite asymptotic variance with the order $O(N^{-1})$ given that ESS is away from $0$. Here, we focus on Algorithm \hyperlink{1}{3}. Since the stability of the conditional gradient ascent update is determined by the variance of $\widehat{\nabla}\mathrm{log}p_{\theta}(y_t|y_{0:t-1})$, below the asymptotic variance of estimator using $\{(x_{0:t}^{(i)},w_t^{(i)})\}_{i=1}^N$ is given to highlight the impact of the time-varying $\theta_t$ and justify that it can be controlled using the averaged ESSs in the renewing step. 

Suppose multinomial resampling is used at every step, and the renewal does not occur until time t. Denote $p_{\theta_{k}}(\cdot|y_{0:t})$ by $\pi_{\theta_{k}}^{t}(\cdot)$. 
\begin{theorem}
\label{main_thm}
For a test function $f$ and the initialisation of particles satisfying certain regularity conditions(see assumption B in the supplementary materials), conditional on $\theta_1,\cdots,\theta_t$, as $N\rightarrow\infty$ we have:
\vspace{-2pt}
\begin{equation*}
{\small
\begin{aligned}
&\sqrt{N}\left(\frac{\Sigma_{i=1}^{N}w_{t}(x_{0:t}^{(i)})f(x_{0:t}^{(i)})}{\Sigma_{i=1}^{N}w_{t}(x_{0:t}^{(i)})}-\mathbb{E}(\pi_{\theta_{t}}^{t}(f))\right)\xrightarrow{\text{D}}N(0,V_{t}(f)),
\end{aligned}
}
\end{equation*}
\vspace{-2pt}
\begin{equation*}
\begin{aligned}
&\text{where} \;\; V_{t}(f)=\int \frac{\pi_{\theta_{t}}^{t}(x_{0})^{2}}{q_0(x_{0})} \cdot\\
&\left(\int f(x_{0:t})\pi_{\theta_{t}}^{t}(x_{1:t}|x_{0})dx_{1:t}-\mathbb{E}_{\pi_{\theta_{t}}^{t}}(f(x_{0:t}))\right)^{2}dx_{0}\\
&+\Sigma_{k=1}^{t}\int \frac{\pi_{\theta_{t}}^{t}(x_{0:k})^{2}}{q_k(x_{k}|x_{0:k-1})\pi_{\theta_{k}}^{k-1}(x_{0:k-1})}\cdot\\
&\left(\int f(x_{0:t})\pi_{\theta_{t}}^{t}(x_{k+1:t}|x_{0:k})dx_{k+1:t}-\mathbb{E}_{\pi_{\theta_{t}}^{t}}(f(x_{0:t}))\right)^{2}dx_{0:k}. 
\end{aligned}
\end{equation*}

As a result, in Algorithm \hyperlink{1}{3}, as $N\rightarrow\infty$, 
$$
\widehat{\nabla}\log p_\theta(y_t|y_{0: (t-1)})\xrightarrow{\text{P}} \nabla\log p_\theta(y_t|y_{0: (t-1)}).
$$
\end{theorem}
The complete statement is presented in the supplements. The theorem extends the standard asymptotic variance result \citep{chopin2004central} in that each importance sampling variance term contains the time-indexed parameter in both the target densities $\pi_{\theta_{t}}^{t}(x_{0:k})$ and the proposal densities $q_k(x_{k}|x_{0:k-1})\pi_{\theta_{k}}^{k-1}(x_{0:k-1})$ over $k=0,\cdots,t$. The variance increases as $\theta_{t}$ increasingly differs from $\theta_{k}$ with $k$ fixed as $t$ increases, since the importance weight in the $k$th term is proportional to 
\begin{equation}
\frac{p_{\theta_k}(x_k, y_k | x_{0: k-1}, y_{0: k-1})}{q_k(x_k | x_{0: k-1})}\cdot\frac{p_{\theta_t}(x_{0: k}, y_{0: t})}{p_{\theta_k}(x_{0: k},y_{0: k})}.  \label{eq:asy_var_term}  
\end{equation}
The increase in variance comes from the second term. 

\section{NUMERICAL EXPERIMENTS}
This section presents a comparative analysis of the performance of \textit{adaptGA-PIS} and \textit{semiGA-PIS} using the state-of-the-art (SoTA) algorithms from \cite{poyiadjis2011particle}, denoted by P-offline and P-online, as the benchmarks. We also compare their performance with \textit{Naive SGA} and \textit{Fisher SGA}, which are the SGA algorithms using particles from the vanilla SMC in Algorithm \ref{onlineMLE}. The difference is that the former estimates the score function using finite difference ignoring the discontinuity of the SMC likelihood, and the latter uses the Fisher's identity \eqref{Fisher_ident}. The semiGA-PIS algorithm is compared with Algorithm \ref{onlineMLE}, named \textit{vanilla onlineGA}, to illustrate its robustness. The algorithms are tested on three models: the noisy auto-regressive(AR) model, the stochastic volatility (SV) model and the Poisson auto-regressive(PAR) model. 

The root mean square errors (RMSE) with respect to the MLE are reported. In the plots, the markers indicate the times at which values of the estimated parameters are output. For all algorithms, the multinomial resampling is performed at every step, i.e. $r_2=1$. The step sizes of the online algorithm are $\gamma_{t}$ multiplied by the data length $T$ so that the gradient updates in all algorithms have a similar scale. Relaxing $r_1$ to some reasonable value such as $0.5$ can further reduce the computational cost of Algorithm \ref{onlineAIS}. The algorithms are compared with the same CPU times which are measured using the $time.perf\_counter()$ command in Python. The choices of learning rates in all models based on the guidelines from \citet{spall1998implement}: $\gamma_{t}$ is in the form of $\frac{c_{1}}{(A+t)^{\alpha}}$ with $\alpha=1$. The joint choice of $c_{1}, A$  are based on the magnitude of the gradient and the scale of the parameter. The learning rates are reasonable and fair in the sense that their scales are sufficient for multiple algorithms to converge. More details including raw data and Python code can be found in the \hyperlink{2}{supplementary materials}.

\subsection{Auto-regressive Model of Order One with Noise}

Consider the univariate AR(1) model with added normal noise, given below,
\begin{equation}
\label{ar1}
    X_{t+1}= \phi X_{t}+\sigma_x\eta_{t},\;
     Y_{t}= X_{t}+\sigma_y\xi_{t},\;\;\; t=0,\dots, T,
\end{equation}
where $\eta_{t},\xi_{t}$ are independent standard Gaussian noise, $X_0$ has mean $0$ and follows the stationary distribution of $X_t$. The parameter of interest is $\theta \triangleq(\phi,\sigma_{x},\sigma_{y})$ and its MLE can be computed by the Kalman filter. The data length is chosen to be a moderately large value $T=10000$ which is challenging to using the importance sampling weights. The particle size $N=1000$ for all algorithms, and for Algorithm \ref{onlineAIS} we set $r_{1}=0.5$. 


First, we compare all algorithms except Algorithm \ref{onlineMLE}. The results are presented below:
\begin{table}[h]
  \caption{\small{RMSE ratios for the AR(1) experiment from $20$ replications averaged over all parameters.  For all algorithms $N=1000$ except P-offline and P-online which costs $O(N^{2})$, while for the two we choose $N=30$ due to the computational cost restriction. The initial parameters $\theta_{0}=(0.5,0.5,0.7)$,  $r=0.2$ for Algorithm \ref{offline_algo} and $r_{1}=0.5$ for Algorithm \ref{onlineAIS}. RMSEs are calculated with respect to $\theta_{MLE}=(0.66825, 0.73900, 0.95750)$ estimated by the Kalman filter. The CPU time is 130 seconds.}}
  \label{ex1_2}
  
  \centering
  \begin{tabular}{llll}
    \toprule
    Algorithms     & $\phi$  &$\sigma_{x}$ &$\sigma_{y}$ \\
    \midrule
    P-offline $O(N^{2})$ &4.106  &1.514 &1.945\\
    P-online $O(N^{2})$ &1.000 &1.000 &1.000\\
    Naive SGA &6.900  &6.393 &10.13  \\
    Fisher SGA     &5.556 &5.522 &9.29 \\
    AdaptGA-PIS &1.461 &1.539 &1.830\\
    semiGA-PIS &0.568 &0.595 &0.612  \\
    \bottomrule
  \end{tabular}
\end{table}

The table shows that our proposed semi-online algorithm has the lowest RMSE among all algorithms. The algorithm performed 393 renewing steps in total over the 20 replications. The P-online has the second lowest RMSE. Our proposed adaptGA-PIS algorithm performs the best among all offline algorithms, maintaining a lower RMSE with a given computational budget, followed by P-offline. The naive SGA performs the worst due to the discontinuity of the SMC likelihood $\widetilde{l}(\theta)$, and although the Fisher SGA solves this issue, it does not have much improvement.

\begin{figure}[h]
  \centering

\includegraphics[width=0.32\textwidth]{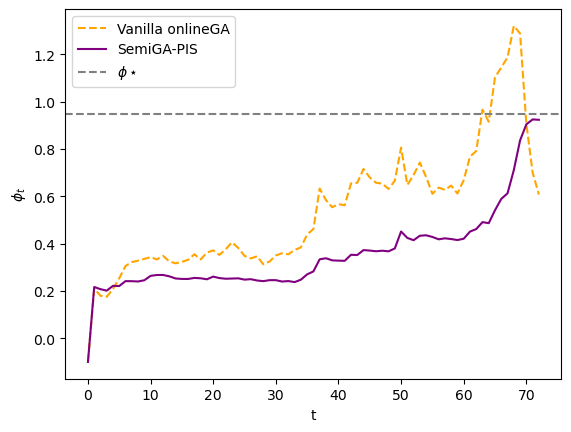}
\includegraphics[width=0.32\textwidth]{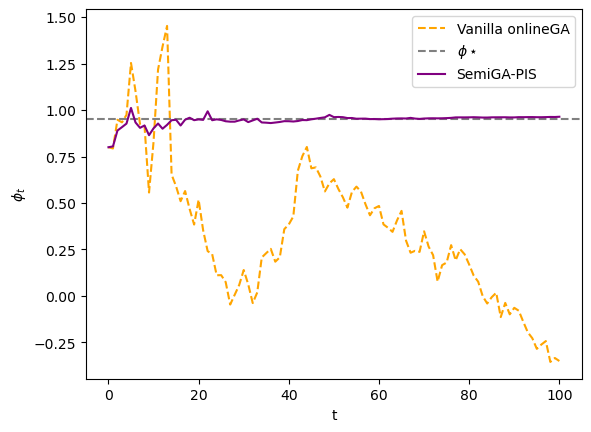}
  \caption{\small{Trajectory plots for one replication of model \eqref{ar1} (Left) and model \eqref{ar1_changed} (Right), with initial values $-0.1$ and $0.8$ respectively. Both models have $\phi_\star=0.95$ and $\sigma_x=0.5$ and $\sigma_y=0.5$.}}
\label{robust}
\end{figure}

Second it is shown that Algorithm \ref{onlineAIS} is more robust than Algorithm \ref{onlineMLE}. The following model is tested where a degenerating trend term is added to $Y_t$:
\begin{equation}
\label{ar1_changed}
    X_{t+1}= \phi X_{t}+\sigma_x\eta_{t},\;
     Y_{t}= 3\phi^t+X_{t}+\sigma_y\xi_{t},\;\;\; t=0,\dots, T.
\end{equation}
    In model \eqref{ar1_changed}, the early observations are highly informative about $\theta$ by the trend term. The result is in Figure \ref{robust}.

\setlength{\textfloatsep}{13pt}

In both plots, Algorithm \ref{onlineMLE} detects the correct gradient in early updates but eventually fails due to the bias in the particle trajectory. In the right plot although the initial value is fairly close to $\phi_*$, it fails significantly because the wrong $\phi$ also appears in the observation equation and is more misleading. Furthermore, the experiment is run for $20$ replications. We counted the `failure' of Algorithm \ref{onlineMLE}, defined by that $\theta_{end} \notin (0.6,1.3)$ or is a $nan$ value. Among the 20 replications, Algorithm \ref{onlineMLE} has 5 failures for model \eqref{ar1} and 16 for \eqref{ar1_changed}, and Algorithm \ref{onlineAIS} has no failure. Similarly, for model \eqref{ar1}, the RMSE of Algorithm \ref{onlineAIS} over Algorithm \ref{onlineMLE} at the end of the iterations is $0.1129/0.2544=0.44$, and that for model \eqref{ar1_changed} is $0.0140/0.3250=0.043$ due to the many failures of Algorithm \ref{onlineMLE}. More results are in the supplementary materials. The results above show that the negative impact of the inconsistent conditional score estimator in Algorithm \ref{onlineMLE} can be significant when the parameter value trajectory drifts away from $\theta^{\star}$. In practice, it is preferable to use batch data to perform offline algorithms for initialisation, but the pilot stage introduces additional computational cost, and for high-dimensional parameters, the initialisation may still be far from the MLE.

\subsection{Stochastic Volatility Model}
The SV model is a popular class of models to capture the stylised fact of volatility clustering in financial time series. Consider the following univariate SV model from   \cite{sandmann1998estimation}:
\begin{equation}
\label{sv}
\begin{aligned}   
    & X_{t+1}= \phi X_{t}+\sigma_x\eta_{t},\; t=0,\dots, T-1,\\
    & Y_{t}= \sigma_{y}e^{\frac{X_{t}}{2}}\xi_{t},\; t=0,\dots, T,\\
\end{aligned}
\end{equation}
where $X_0$ has mean $0$ and $\eta_{t},\xi_{t}$ follow $N(0,1)$. The parameter to estimate is $\theta=(\phi,\sigma_{x},\sigma_{y})$. Since the SV model is non-linear and non-Gaussian, the Kalman filter can not be applied. The RMSE is calculated with respect to the estimated value $\hat{\theta}_{MLE}=(0.896, 0.399, 0.243)$, which is obtained by running Fisher SGA with large values of $N$ and iteration numbers initialised from the true parameter value. The RMSE ratios are shown in Table \ref{ex2}.


\begin{table}[h]
  \caption{\small{RMSE ratios for the SV experiment from $50$ replications averaged over all parameters.The true parameter $\theta^{\star}=(0.9,0.40,0.25)$ and $T=10000$. For all algorithms $N=1000$ except P-offline and P-online which costs $O(N^{2})$, while for the two we choose $N=30$ due to the computational cost restriction. The initial parameters $\theta_{0}=(0.7,0.25,0.40)$,  $r=0.4$ for Algorithm \ref{offline_algo} and $r_{1}=0.6$ for Algorithm \ref{onlineAIS}. The semi-online algorithm performs 6665 renewing steps over the 50 replications. The CPU time is 3000 seconds. The results using $N=400$ (and $N=20$ for $O(N^2)$ the algorithms) are similar}}
  \label{ex2}
  \centering
  \begin{tabular}{llll}
    \toprule
    Algorithms     & $\phi$  &$\sigma_{x}$ &$\sigma_{y}$ \\
    \midrule
    P-offline $O(N^{2})$ &2.377  &2.043 &0.222\\
    P-online $O(N^{2})$ &1.000 &1.000 &1.000\\
    Naive SGA &4.535 &2.531 &1.989 \\
    Fisher SGA     &3.908 &1.681  &0.899 \\
    AdaptGA-PIS &0.165 &0.798 &0.039\\
    SemiGA-PIS &0.226 &0.268 &0.064 \\
    \bottomrule
  \end{tabular}
\end{table}

Table \ref{ex2} shows that both new algorithms converge faster than the SoTA online algorithm in all parameters.  The AdaptGA-PIS algorithm has the lowest RMSE in $\phi$ and $\sigma_{y}$, yet its estimation for $\sigma_{x}$ is considerably worse than the SemiGA-PIS algorithms. The Naive SGA performs the worst, with the Fisher SGA improves moderately. The P-offline algorithm performs better than the Fisher SGA and Naive SGA overall with the $O(N^2)$ cost. 

It is known that $\phi$ is difficult to identify when the other parameters have certain values \citep[Chapter 14]{chopin2020introduction}. Below the behaviours of our algorithms in regions with flat likelihood surfaces stated back in Section 3 are demonstrated by the figures below.
We noticed that indeed Algorithm \ref{offline_algo} required fewer SMC runs to regenerate particles when moving on the flat surface, and when there are significant changes in likelihood the algorithm detects it and ensures the following iterations are in the right direction. Algorithm \ref{onlineAIS} has a similar behaviour and we can see clearly from the second plot that between the second renewing step and the third renewing step, the parameter is not moving in the right direction since the gradient estimates are based on the particles generated under the second renewing step. But the algorithm detected the issue and after the third renewing step the parameter started to update in the right gradient. This is another example where our algorithm is more robust than Algorithm \ref{onlineMLE}, where there is no renewing step and in early iterations the gradient updates may suffer from the bad quality of the particles.

Similar to model \eqref{ar1_changed}, a model with a degenerating trend added to the mean volatility is tested. The RMSE of the semi-online algorithm significantly outperforms all others by up to one order of magnitude. Again it shows the benefit of the exact online updates when early observations are highly informative. Specific results are in the supplementary materials.  

\subsection{Poisson Auto-regressive Model}
Consider the real-world time series of $168$ monthly counts of poliomyelitis in the United States from January 1970 to December 1983, introduced by   \cite{zeger1988regression}. The dataset is well studied using the Poisson Auto-regressive model of order 1(PAR(1)) with a deterministic trend function nonlinear in $t$, where there are eight unknown parameters $\theta=(\mu_{1},\mu_{2},\mu_{3},\mu_{4},\mu_{5},\mu_{6},\phi,\sigma_{x})$ \citep{langrock2011some}.
Our implementation of the SGA algorithm with $2000$ iterations gives parameter estimates consistent with the literature and is used to calculate the RMSEs. Since P-offline and P-online have no benefits for short time series given the $O(N^2)$ cost, they are not compared here. Occasionally the naive SGA updates move to the wrong direction due to the discontinuity of $\widetilde{l}(\theta)$, and give invalid values for the parameters. The next least feasible random seed is used in such cases. The semi-online algorithm is used as a benchmark and the results are in Table \ref{rmseratios}.

\begin{table}[h]
  \caption{\small{RMSE ratios for the PAR(1) experiment from $20$ replications averaged over all parameters. For all algorithms $N=3000$, the initial parameters $\theta_{0}=(0.4, -3.8, 0.2,-.4,0.5,-0.1,0.7,\sqrt{0.4})$,  $r=0.6$ for Algorithm \ref{offline_algo} and $r_{1}=0.6$ for Algorithm \ref{onlineAIS}. The CPU time is 900 seconds.}}
  \label{rmseratios}
  
  \centering
  \begin{tabular}{lllll}
    \toprule
    Algorithms     & $\mu_{1}$    &$\mu_{2}$  &$\mu_{3}$  &$\mu_{4}$ \\
    \midrule
    Naive SGA & 3.829 &1.242 &1.266 &1.590  \\
    Fisher SGA     & 0.519 &0.944 &0.960&1.196 \\
    AdaptGA-PIS &0.674  &0.8645 &0.276 &0.455 \\
    SemiGA-PIS &1.000 &1.000 &1.000 &1.000  \\
    \bottomrule
    \toprule
    Algorithms   &$\mu_{5}$  &$\mu_{6}$  &$\phi$  &$\sigma$    \\
    \midrule
    Naive SGA &2.1955 &2.4885 &1.7565 &1.240 \\
     Fisher SGA  &1.462 &1.646 &1.337 &0.745 \\
     AdaptGA-PIS &0.258 &0.383 &0.6165 &0.491 \\
     SemiGA-PIS &1.000 &1.000 &1.000 &1.000 \\
    \bottomrule
  \end{tabular}
\end{table}
The AdaptGA-PIS algorithm shows superior performance by having a low RMSE in all parameters. The SemiGA-PIS algorithm is inferior here due to the short data size ($T=168$). This can be seen by that with $O(TN)$ computational cost, the semi-online algorithm moves $T$ steps while the offline algorithms move 1 step, so a larger $T$ may give more advantage of the semi-online algorithm over the offline algorithms. 
\section{CONCLUSIONS}
Two SMC algorithms are proposed to utilize the importance weight $a_{\theta_0}(\theta,x_{0:t})$ for gradient-based likelihood optimization. In offline gradient ascent the computational efficiency can be significantly improved by recycling generated particles for multiple updates where the approximation accuracy is maintained by thresholding the ESS. The semi-online algorithm performs parameter updates within the SMC steps and re-targets the particle mass using $a_{\theta_t}(\theta_{t+1},x_{0:t})$ to achieve consistent estimation. The benefits of controlled particle quality in offline updates and the efficiency of online updates are combined by regenerating the particles when the averaged ESSs is low which approximately controls the asymptotic variance of the conditional score estimator. Both algorithms are robust to the choice of step size and flat likelihood surface, and show competitive performance to SoTA algorithms in numerical studies with long time series. 

This work offers several interesting avenues for further research. One is to modify the structure of the semi-online algorithm with different control measures for the renewing step, which may be more computationally efficient than the ESS. Another is to use the semi-online algorithm for deterministic optimisation of criterion function which can be expressed in the form of a state-space model.

\paragraph{LIMITATIONS}
The two new algorithms are suitable for models where simulating particles is more expensive than evaluating the joint log-likelihood. For example, for each model in the numerical study, the joint likelihood depends on the latent states through a $d$-dimensional summary statics with $d$ fixed as $t$ increases, hence only a $d\times N$ matrix needs to be stored and updated to evaluate the ESSs as each algorithm iterates. This feature exists for a class of state-space model and is often utilized for MCMC moves in SMC algorithms \cite[]{fearnhead2002markov,kantas2015particle}. For other models, e.g. a $t\times N$ matrix needs to be stored and updated as $t$ increases, the performance of two algorithms may drop and the memory requirement needs to be considered. The tuning of the ESS thresholds $r_{1},r_{2}$ in the online algorithm is model-dependent and may require extra work. Theorem 1 considers a simplified case that the central limit theorem at time t is conditional on the trajectory of $\theta$. Since the trajectory depends on the particle path, the unconditional result is more involved than Theorem 1.

{
\small

\bibliography{references}

}

\newpage
\onecolumn
\aistatstitle{Parameter Estimation in State Space Models Using Particle Importance
Sampling: Supplementary Materials}
\appendix
\section{THE SMC FRAMEWORK}
\hypertarget{3}
We have introduced the vanilla SMC as a case in Algorithm \ref{onlineMLE}, here we provide an alternative expression which may be clearer for some readers: 
\begin{algorithm}[h]
  \caption*{Vanilla sequential Monte Carlo}
  \begin{algorithmic}
   \State
   Assume that $f,g,\theta$, $y_{0:T}$ and the number of particles needed $N$ are given, and a proposal distribution for generating new latent variables $q_{\theta}(x_{t}|x_{0:(t-1)})$ is available. For $t=0$, the proposal distribution is $q_{\theta}(x_{0})$), and we set all $\widetilde{w}_{-1}^{(i)}=1$, $i=1,\cdots,N$\\
\noindent \textbf{For} $t = 0$ to $T$:
\begin{adjustwidth}{1cm}{0cm}
\textbf{Propagation:} Use the proposal distribution $q_{\theta}(x_{t}^{(i)}|\tilde{x}_{0:(t-1)}^{(i)})$ to sample the new latent variable, and let $x_{0:t}^{(i)}=(\tilde{x}_{0:(t-1)}^{(i)},x_{t}^{(i)})$, $i = 1$ to $N$.\\
\noindent\textbf{Reweightning:}
Compute $u_{t}^{(i)}$, the $incremental \; weight$ for the particle, defined as:\\
$u_{t}^{(i)}\triangleq \frac{g_{\theta}(y_{t}|x_{t}) f_{\theta}(x_{t}^{(i)}|\tilde{x}_{t-1}^{(i)})}{q_{\theta}(x_{t}^{(i)}|\tilde{x}_{0:(t-1)}^{(i)})}$, $i = 1$ to $N$.\\
Set the particle $x_{0:t}^{(i)}$ with weight $w_{t}^{i}$ where $w_{t}^{(i)}\triangleq u_{t}^{(i)}\tilde{w}_{t-1}^{(i)}$\\
\noindent\textbf{Resampling:}
If the condition for resampling is satisfied, resample $\{x_{0:t}^{(i)}\}_{i=1}^{N}$ according to corresponding weights $\{w_{t}^{(i)}\}_{i=1}^{N}$ to obtain $\{\tilde{x}_{0:t}^{(i)}\}_{i=1}^{N}$ and set all $\{\tilde{w}_{t}^{(i)}\}_{i=1}^{N}$ equals 1; if the condition for resampling is not satisfied, set $\{\tilde{x}_{0:t}^{(i)}\}_{i=1}^{N}=\{x_{0:t}^{(i)}\}_{i=1}^{N}$ and $\{\tilde{w}_{t}^{(i)}\}_{i=1}^{N}=\{w_{t}^{(i)}\}_{i=1}^{N}$ 
\end{adjustwidth}
\textbf{End For}
  \end{algorithmic}
\end{algorithm}
At each time $t$, the weighted particles$\{(\tilde{x}_{0:t}^{(I)},\tilde{w}_{t}^{(i)})\}_{i=1}^{N}$ is an approximation of the samples from distribution $p_{\theta}(x_{0:t}|y_{0:t})$. Therefore, for any function $h(x_{0:t})$, we could estimate the expectation $E_{\theta}(h(x_{0:t})|y_{0:t})$ by $\frac{\Sigma_{i=1}^{N}h(\tilde{x}_{0:t}^{(i)})\tilde{w}_{t}^{(i)}}{\Sigma_{i=1}^{N}h(\tilde{w}_{t}^{(i)})}$. Suppose that the resampling happens at time $t_{j}, j=1,\cdots,k $, and we define $t_{k+1}\triangleq T$. The estimate of the likelihood is given by: $\tilde{p}_{\theta}(y_{0:T})=\Pi_{j=1}^{k+1} (N^{-1}\Sigma_{i=1}^{N}w_{t_{j}}^{(i)})$. It was shown that both $(\Sigma_{i=1}^{N}w_{t}^{(i)})/N$ and $\tilde{p}_{\theta}(y_{0:T})$ are unbiased estimators and are asymptotic normal with convergence rate $\sqrt{N}$ as $N \to \infty $\citep{kantas2015particle}. We can then estimate the log-likelihood by the particles :
\begin{equation}
\label{llhd es}
   \tilde{l}(\theta)\triangleq \mathrm{log}(\tilde{p}_{\theta}(y_{0:T}))=\Sigma_{j=1}^{k+1} \mathrm{log}(N^{-1}\Sigma_{i=1}^{N}w_{t_{j}}^{(i)}), 
\end{equation}
which is also unbiased and asymptotic normal with convergence rate $\sqrt{N}$ as $N \to \infty$. 

\section{MORE ON MONTE CARLO LIKELIHOOD FUNCTION}
We mentioned in section 2 that in the smoothed likelihood approximation in equation (\ref{lre}), the importance sampling estimator is under the true distribution $p_{\theta_{true}}(x_{0:T}|y_{0:T})$ and the proposal distribution $p_{\theta_{last}}(x_{0:T}|y_{0:T})$. Where $\theta_{last}$ is the last parameter to generate the particles. We have made this claim, and we would like to briefly justify it using the toy example below.
\paragraph{Example} To illustrate the levels of accuracy of the $\hat{l}_{\theta_{0}}(\theta)$ approximation, consider estimating the log-likelihood function of a first-order auto-regressive model (AR(1)). Following the notation in (\ref{ar1}) and assume that $\sigma_{x},sigma_{y}$ are known, so the parameter of interest $\theta \triangleq \phi$. The true parameter value is $\theta_{true}=0.7$, $N=1000, T=5, \mu_{0}=0,\Sigma_{0}=10,V_{t}\sim N(0,1),W_{t}\sim N(0,0.5^{2})$. The usual random seed $s=1$ is used. In this case, we could use the Kalman filter to obtain the true likelihood curve $l(\theta)$, and we also use independent SMC samples to pointwise approximate the likelihood values, which means numerous sets of particles are needed. We then use three different simulation parameters $\theta_{0}$, each generating a set of particles, and use the likelihood ratio approximation to produce a smoothed likelihood curve estimate $\Tilde{l}_{\theta_{0}}(\theta)$ for $l(\theta)$. The results are shown below. 

\begin{figure}[h]
  \center
\includegraphics[width=0.7\textwidth]{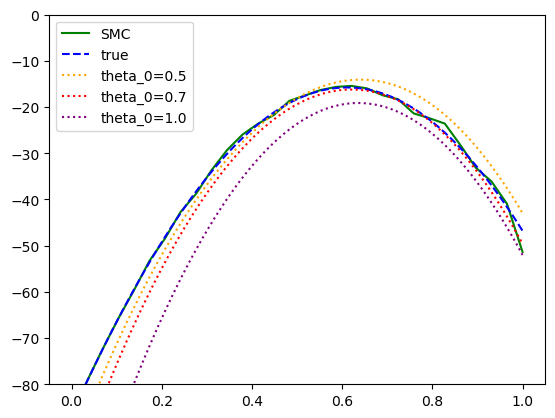}\\\	
  \caption{Likelihood approximation of an AR(1) model using independent SMC estimate and smoothed likelihood ratio estimates $\hat{l}_{\theta_{0}}(\theta)$ under different $\theta_{0}$}
\end{figure}
We can see that the SMC approximation is close to the true likelihood curve, but several sets of particles are required and the approximation is not continuous due to the Monte Carlo noise. When $\theta_{0}=0.7=\theta_{true}$, i.e. the true parameter value is used for the simulation, $\hat{l}_{\theta_{true}}(\theta)$ gives a very nice approximation to $l(\theta)$ using only one set of particles. If
$\theta_{0}$ is far from the true parameter value, i.e. for $\theta_{0}=0.5,1$ both methods have good approximation accuracy for $\theta$ around the corresponding $\theta_{0}$, but $\hat{l}_{\theta_{0}}$ becomes increasingly biased as $\theta$ is far from $\theta_{0}$.

\section{ASSUMPTIONS and PROOF of Theorem 1}
This section provides proof for Theorem \ref{main_thm} in section 3 of the main text. We follow the framework of \citet{douc2008limit} but the information here should be sufficient for justifying our theories.

\textbf{Definition 1} 
Let $L^{1}(\mathcal{X},\mu) \triangleq \{f:\mathcal{X}\mapsto \mathbb{R}|\;\int |f(x)|\mu(x)dx\}<\infty$, The set of weighted samples $\{x_{0:t}^{(i)},w_t^{(i)}\}_{i=1}^{N}$ is said to be consistent for the probability measure $\mu$ and the set $\mathcal{C}\subseteq L^{1}(\mathcal{X},\mu)$ if for any $f\in \mathcal{C}$, as $N\to \infty$:
\begin{equation}
\label{consistentmean}
 \Omega_{t}^{-1}\Sigma_{i=1}^{N}w_{t}^{(i)}f(x_{0:t}^{(i)}) \xrightarrow{\text{P}} E_{\mu}(f(x_{0:t})),
\end{equation}
\begin{equation}
\label{consistentmax}
 \Omega_{t}^{-1} \max_{1\leq i \leq N} w_{t}^{(i)} \xrightarrow{\text{P}} 0,  \text{where}\; \Omega_{t}\triangleq \Sigma_{i=1}^{N}w_{t}^{(i)}.
\end{equation}

\textbf{Definition 2} 
The set of weighted samples $\{x_{0:t}^{(i)},w_t^{(i)}\}_{i=1}^{N}$ is said to be asymptotically normal for $\{\mu,\gamma,A,W,\sigma\}$ if:
\begin{equation}
\label{asympmean}
 \sqrt{N}\Omega_{t}^{-1}\Sigma_{i=1}^{N}w_{t}^{(i)}(f(x_{0:t}^{(i)})-E_{\mu}(f(x_{0:t})) \xrightarrow{\text{D}} N(0,\sigma^2(f)), \forall f \in A,
\end{equation}
\begin{equation}
\label{asympvar}
 N\Omega_{t}^{-2}\Sigma_{i=1}^{N}(w_{t}^{(i)})^{2}f(x_{0:t}^{(i)}) \xrightarrow{\text{P}} \gamma(f), \forall f \in W,
\end{equation}
\begin{equation}
\label{asympmax}
 \sqrt{N}\Omega_{t}^{-1} \max_{1\leq i \leq N} w_{t}^{(i)} \xrightarrow{\text{P}} 0.
\end{equation}
\textbf{Definition 3(Proper Set)}
We say that the set $\mathcal{C}$ is proper if the following conditions hold:\begin{equation}
\label{properset}
\begin{aligned}
&(i) Let\; a,b\in \mathbb{R} and f,g\in \mathcal{C}\; then\; af+bg\in \mathcal{C}.\\ 
&(ii)Let\;  f\in \mathcal{C}\; and\; g\;is \; measurable\;with\; |g|<|f|\; then\; g\in\mathcal{C}.\\
&(iii) Any\;c\in \mathbb{R}\; the\;constant\;function\;f\equiv c \in \mathcal{C}. 
\end{aligned}
\end{equation}
We assume that the trace of our parameters satisfies regularity conditions:

\textbf{Assumption A} we assume that the trace of our gradient ascend algorithm lies in a set $\Theta$ and it is proper in the sense that, given any $x_{0:t}, \{\theta_{i}\}_{i=0}^{t+1}$, we have that $\forall 0\leq i,j \leq t+1 :$
\begin{equation}
\label{assumptionbound}
\begin{aligned}
&\frac{p_{\theta_{i}}(x_{0:t}|y_{0:t})}{p_{\theta_{j}}(x_{0:t}|y_{0:t})}\in [l_{1}(x_{0:t},y_{0:t}),u_{1}(x_{0:t},y_{0:t})],\\
&\frac{p_{\theta_{i}}(x_{0:t},y_{0:t})}{p_{\theta_{j}}(x_{0:t},y_{0:t})}\in [l_{2}(x_{0:t},y_{0:t}),u_{2}(x_{0:t},y_{0:t})] ,
\end{aligned}    
\end{equation}
where the bounds $l_{1}(x_{0:t},y_{0:t}),u_{1}(x_{0:t},y_{0:t}),l_{2}(x_{0:t},y_{0:t}),u_{2}(x_{0:t},y_{0:t})$ are positive. This is not a strong assumption as when samples are given, it may only fail when $\{\theta_{i}\}_{i=0}^{t+1}$ contain value(s)$\theta_{i}$ that makes $p_{\theta_{i}}(x_{0:t}|y_{0:t})$ or $p_{\theta_{i}}(x_{0:t},y_{0:t})$ arbitrarily close to 0 or $\infty$, which is unlikely to occur and will particularly leads to failure of the algorithm by producing $nan$ values, as mentioned in section \hyperlink{nan}{5.1.1}. Henceforth, conditioned on a set of weighted samples and observations, we may drop the $x_{0:t},y_{0:t}$ from $l$ and $u$.

\textbf{Assumption B}
Under assumption A, denote $\pi_{\theta}^{k}\triangleq p_{\theta}(x_{0:k}|y_{0:k})$. Suppose that $A_{0}$ and $W_{0}$ are proper sets and we recursively define:
\begin{equation}
\label{recursive}
\begin{aligned}
&A_{k}\triangleq \{f\in L^{2}(\mathcal{X},\pi^{k}_{\theta}),E_{q_{k}}[\frac{\pi^{k}_{\theta}}{q_{k}\pi^{k-1}_{\theta}}f\;|x_{0:k-1}]\in A_{k-1}, E_{q_{k}}[(\frac{\pi^{k}_{\theta}}{q_{k}\pi^{k-1}_{\theta}}f)^2\;|x_{0:k-1}]\in W_{k-1}, \;\forall \theta \in \Theta \},\\ 
&W_{k}\triangleq \{f\in L^{1}(\mathcal{X},\pi_{\theta}^{k}), E_{q_{k}}[(\frac{\pi^{k}_{\theta}}{q_{k}\pi^{k-1}_{\theta}}f)^2\;|x_{0:k-1}]\in W_{k-1},\; \forall \theta \in \Theta \}; \;\;\;\;\;\;\; k\geq 1.
\end{aligned}
\end{equation}
We also suppose that the initialisation $\{x_{0}^{(i)}\}_{i=0}^{N}$ are consistent for $\{p_{\theta_{0}}(x_{0}|y_{0}),L^{1}(\mathcal{X},p_{\theta_{0}}(x_{0}|y_{0})\}$ and asymptotically normal for $\{p_{\theta_{0}}(x_{0}|y_{0}),p_{\theta_{0}}(x_{0}|y_{0}),A_{0},W_{0},\sqrt{Var_{q_{0}}(f)}\}$.

\begin{lemma}
Assume that the weighted samples $\{x_{0:t}^{(i)},w_t^{(i)}\}_{i=1}^{N}$ is consistent for $\{p_{\theta_{t}}(x_{0:t}|y_{0:t}), \mathcal{C} \}$, and denote $a_{\theta}(\theta_{t+1},x_{0:t},y_{0:t})$ by $a_{t}^{t+1}$, then after the retargeting step $\tilde{w}_{t}^{(i)}\triangleq a_{t}^{t+1}w_{t}^{(i)}$, the new weighted samples $\{x_{0:t}^{(i)},\tilde{w}_t^{(i)}\}_{i=1}^{N}$ is consistent for $\{p_{\theta_{t+1}}(x_{0:t}|y_{0:t}), \mathcal{C} \}$.
\end{lemma}

\begin{lemma}
\label{retargetasymp}
Assume that the weighted samples $\{x_{0:t}^{(i)},w_t^{(i)}\}_{i=1}^{N}$ is asymptotically normal for $\{p_{\theta_{t}}(x_{0:t}|y_{0:t}),\gamma,A_{t},W_{t},\sigma\}$, then after the retargeting step $\tilde{w}_{t}^{(i)}\triangleq a_{t}^{t+1}w_{t}^{(i)}$, the new weighted samples $\{x_{0:t}^{(i)},\tilde{w}_t^{(i)}\}_{i=1}^{N}$ is asymptotically normal for $\{p_{\theta_{t+1}}(x_{0:t}|y_{0:t}),\tilde{\gamma},A_{t},W_{t},\Tilde{\sigma}\}$.
And $\tilde{\gamma}(f)=(\frac{p_{\theta_{t}}(y_{0:t})}{p_{\theta_{t+1}}(y_{0:t})})^{2}{\gamma}((a_{t}^{t+1})^{2}f),\Tilde{\sigma}^{2}(f)=(\frac{p_{\theta_{t}}(y_{0:t})}{p_{\theta_{t+1}}(y_{0:t})})^{2}{\sigma}^{2}(a_{t}^{t+1}f).$
\end{lemma}
Combining lemma 1,2 with the results from \citet{chopin2004central}, we propose the following theorem for the SemiGA-PIS algorithm with varying parameters.
\begin{theorem}
\label{AIStheorem}   
Assume that we are in the context of Algorithm 3 in which the multinomial resampling is employed at every step, while renewing does not occur until time t. Denote $p_{\theta_{k}}(y_{0:t})$ by $p_{k}(y_{0:t})$, $p_{\theta_{k}}(\cdot|y_{0:t})$ by $\pi_{k}^{t}$, the proposing distribution $q(x_{t}|\cdots)$ by $q_{t}$. For test function $f$ and the initialisation of particles satisfying assumption B, we let $\tilde{V}_{0}(f)\triangleq \text{Var}_{q_{0}}f$ and recursively define:
\begin{equation}
\label{fourVariances}
\begin{aligned}
&\tilde{V}_{t}(f)=\hat{V}_{t-1}(\mathbb{E}_{q_{t}}(f))+\mathbb{E}_{\pi_{t}^{t-1}}(\text{Var}_{q_{t}}(f)), \;\;t>0 ,\\  
&V_{t}(f)=\tilde{V}_{t}(v_{t}\cdot(f-\mathbb{E}_{\pi_{t}^{t}}(f))),\;\;t\geq 0,\\     
&\bar{V}_{t}(f)=(\frac{p^{t}(y_{0:t})}{p_{t+1}(y_{0:t})})^{2}V_{t}(a_{t}^{t+1}f),\;\;t\geq 0,  \\   
&\hat{V}_{t}(f)=\bar{V}_{t}(f)+\text{Var}_{\pi_{t+1}^{t}}(f),\;\;t\geq 0.     
\end{aligned}
\end{equation}
\end{theorem}
Then we have that:
\begin{equation}
\label{fourconvergences}
\begin{aligned}
&\sqrt{N}(N^{-1}\Sigma_{i=1}^{N}f(x_{0:t}^{(i)})-\mathbb{E}(\pi_{t}^{t-1}q_{t}(f)))\xrightarrow{\text{D}}N(0,\tilde{V}_{t}(f)),\\
&\sqrt{N}(\frac{\Sigma_{i=1}^{N}w_{t}(x_{0:t}^{(i)})f(x_{0:t}^{(i)})}{\Sigma_{i=1}^{N}w_{t}(x_{0:t}^{(i)})}-\mathbb{E}(\pi_{t}^{t}(f)))\xrightarrow{\text{D}}N(0,V_{t}(f)),\\
&\sqrt{N}(\frac{\Sigma_{i=1}^{N}\bar{w}_t(x_{0:t}^{(i)})f(x_{0:t}^{(i)})}{\Sigma_{i=1}^{N}\bar{w}_{t}(x_{0:t}^{(i)})}-\mathbb{E}(\pi_{t+1}^{t}(f)))\xrightarrow{\text{D}}N(0,\bar{V}_{t}(f)),\\
&\sqrt{N}(N^{-1}\Sigma_{i=1}^{N}f(\hat{x}_{0:t}^{(i)})-\mathbb{E}(\pi_{t+1}^{t}(f)))\xrightarrow{\text{D}}N(0,\hat{V}_{t}(f)).\\
\end{aligned}
\end{equation}

Particularly, we specify:
\begin{equation*}
\label{Vexpression}
\begin{aligned}
V_{t}(f)&=\int \frac{(\pi_{t}^{t}(x_{0}))^{2}}{q(x_{0})}(\int f(x_{0:t})\pi_{t}^{t}(x_{1:t}|x_{0})dx_{1:t}-\mathbb{E}_{\pi_{t}^{t}}(f(x_{0:t})))^{2}dx_{0}\\
&+\Sigma_{k=1}^{t}\int \frac{(\pi_{t}^{t}(x_{0:k}))^{2}}{q(x_{k}|x_{0:k-1})\pi_{k}^{k-1}(x_{0:k-1})}(\int f(x_{0:t})\pi_{t}^{t}(x_{k+1:t}|x_{0:k})dx_{k+1:t}-\mathbb{E}_{\pi_{t}^{t}}(f(x_{0:t})))^{2}dx_{0:k},    
\end{aligned}
\end{equation*}
where we let $\int f(x_{0:t})\pi_{t}^{t}(x_{t+1:t}|x_{0:t})dx_{t+1:t}\triangleq f(x_{0:t})$. The expression is also valid for $t=0$ if we just keep the first term and ignore the second term $\Sigma_{k=1}^{0}\cdots$. The expressions for $\hat{V}_{t}(f),\bar{V}_{t}(f),\tilde{V}_{t}(f)$ could thus be easily calculated from $V_{t}(f)$.

The intuition behind this is that combining lemma 1,2 with the result from \cite{chopin2004central} gives the four corresponding variances of the theorem. Then one may simply notice that, although the updated new $\theta_{t+1}$ and thus the retargeting step is based on particles at time $t$, it does not influence the particles between reweighting and retargeting. Thus one may simply combine $\tilde{w}_{t+1}=a_{t}^{t+1}w_{t+1}=a_{t}^{t+1}u_{t+1}\tilde{w}_{t}$ and realise $a_{t}^{t+1}u_{t+1}=\frac{\pi_{t+1}^{t}}{q_{t}\pi_{t}^{t-1}}$, the result then follows as a strict generalisation of \citet{doucet2009tutorial}. An explicit proof using induction was presented after the proofs of Lemma 1 \& 2.


\begin{proof}[Proof of Lemma 1]
We will drop notations $x_{0:t},y_{0:t}$ for simplicity. By assumption (\ref{assumptionbound}) we have that $|a_{\theta}(\theta_{t+1})f|\leq |u_{2}f|$ and $|u_{2}f|\in \mathcal{C}$ by linearity(taking c$\equiv$0 and then use (i)). Thus we have $af\in \mathcal{C}$. Then by applying consistency condition to function $af$ we get:\\
\begin{equation}
\begin{aligned}
\label{meanidentity}
&\Omega_{t}^{-1}\Sigma_{i=1}^{N}\tilde{w}_{t}^{(i)}f(x_{0:t}^{(i)})\\
=&\Omega_{t}^{-1}\Sigma_{i=1}^{N}w_{t}^{(i)}a_{\theta}(\theta_{t+1},x_{0:t}^{(i)},y_{0:t})f(x_{0:t}^{(i)})\\
\xrightarrow{\text{P}}&\int p_{\theta_{t}}(x_{0:t}|y_{0:t})\frac{p_{\theta_{t+1}}(x_{0:t},y_{0:t})}{p_{\theta_{t}}(x_{0:t},y_{0:t})}f\;dx_{0:t}\\
=& \frac{p_{\theta_{t+1}}(y_{0:t})}{p_{\theta_{t}}(y_{0:t})}\int p_{\theta_{t+1}}(x_{0:t}|y_{0:t})f\;dx_{0:t}.
\end{aligned}    
\end{equation}
Particularly, taking $f\equiv1$ we get $\Omega_{t}^{-1}\Sigma_{i=1}^{N}\tilde{w}_{t}^{(i)}=\frac{\Tilde{\Omega}_{t}}{\Omega_{t}}\xrightarrow{\text{P}}\frac{p_{\theta_{t+1}}(y_{0:t})}{p_{\theta_{t}}(y_{0:t})}$. Combining with (\ref{meanidentity}) we get $\tilde{\Omega}_{t}^{-1}\Sigma_{i=1}^{N}\tilde{w}_{t}^{(i)}f(x_{0:t}^{(i)})\xrightarrow{\text{P}}\int p_{\theta_{t+1}}(x_{0:t}|y_{0:t})f\;dx_{0:t}$, as required.

Now we simply notice that $\Tilde{\Omega}^{-1}\max_{1\leq i\leq N}\tilde{w}_{t}^{(i)}\xrightarrow{\text{P}}\frac{p_{\theta_{t}}(y_{0:t})}{p_{\theta_{t+1}}(y_{0:t})}\Omega^{-1}\max_{1\leq i\leq N}\tilde{w}_{t}^{(i)}$ and the fact that $\Omega^{-1}\max_{1\leq i\leq N}\tilde{w}_{t}^{(i)}\leq u_{2}\Omega^{-1}\max_{1\leq i\leq N}w_{t}^{(i)}\xrightarrow{\text{P}}0$, so we have $\Tilde{\Omega}^{-1}\max_{1\leq i\leq N}\tilde{w}_{t}^{(i)}\xrightarrow{\text{P}}0$. 
   
\end{proof}

\begin{proof}[Proof of Lemma 2]

We denote $p_{\theta}(f)\triangleq \int p_{\theta}(x_{0:t}|y_{0:t})f(x_{0:t})dx_{0:t}$. We have that:
\begin{gather*}
\begin{aligned}
&\sqrt{N}\Omega_{t}^{-1}\Sigma_{i=1}^{N} w_{t}^{(i)}\{f(x_{0:t}^{(i)})-p_{\theta_{t}}(f)\}\xrightarrow{\text{D}} N(0,\sigma^{2}(f)),\\
& N\Omega_{t}^{-2}\Sigma_{i=1}^{N} (w_{t}^{(i)})^{2}f(x_{0:t}^{(i)})\xrightarrow{\text{P}}\gamma(f),\\
&\sqrt{N}\Omega^{-1}\max_{1\leq i\leq N}w_{t}^{(i)}\xrightarrow{\text{P}}0.
\end{aligned}  
\end{gather*}
And we first prove that $\sqrt{N}\tilde{\Omega}_{t}^{-1}\Sigma_{i=1}^{N} \tilde{w}_{t}^{(i)}\{f(x_{0:t}^{(i)})-p_{\theta_{t+1}}(f)\}\xrightarrow{\text{D}} N(0,\tilde{\sigma}^{2}(f))$. Notice that $p_{\theta_{t}}(a_{\theta_{t}}(\theta_{t+1})f)=\int p_{\theta_{t}}\frac{p_{\theta_{t+1}}(x_{0:t},y_{0:t})}{p_{\theta_{t}}(x_{0:t},y_{0:t})}f\; dx_{0:t}=\frac{p_{\theta_{t+1}}(y_{0:t})}{p_{\theta_{t}}(y_{0:t})}p_{\theta_{t+1}}(f)$. Combinining this with the previous fact that $\Omega_{t}^{-1}\Sigma_{i=1}^{N}\tilde{w}_{t}^{(i)}=\frac{\Tilde{\Omega}_{t}}{\Omega_{t}}\xrightarrow{\text{P}}\frac{p_{\theta_{t+1}}(y_{0:t})}{p_{\theta_{t}}(y_{0:t})}$ we then get:
\begin{gather*}
\begin{aligned}
&\sqrt{N}\tilde{\Omega}_{t}^{-1}\Sigma_{i=1}^{N} \tilde{w}_{t}^{(i)}\{f(x_{0:t}^{(i)})-p_{\theta_{t+1}}(f)\}\\
=&\sqrt{N}\{\tilde{\Omega}_{t}^{-1}\Sigma_{i=1}^{N} w_{t}^{(i)}a_{\theta_{t}}(\theta_{t+1})f(x_{0:t}^{(i)})\}-\sqrt{N}p_{\theta_{t+1}}(f)\\
=&\sqrt{N}\{\tilde{\Omega}_{t}^{-1}\Sigma_{i=1}^{N} w_{t}^{(i)}a_{\theta_{t}}(\theta_{t+1})f(x_{0:t}^{(i)})\}-\sqrt{N}\frac{p_{\theta_{t}}(y_{0:t})}{p_{\theta_{t+1}}(y_{0:t})}p_{\theta_{t}}(a_{\theta_{t}}(\theta_{t+1})f). 
\;\;\;\;(*)
\end{aligned}
\end{gather*}

By noticing that $\frac{p_{\theta_{t}}(y_{0:t})}{p_{\theta_{t+1}}(y_{0:t})}p_{\theta_{t}}(a_{\theta_{t}}(\theta_{t+1})f)\xrightarrow{\text{P}}\tilde{\Omega}_{t}^{-1}\Sigma_{i=1}^{N}w_{t}^{(i)}p_{\theta_{t}}(a_{\theta_{t}}(\theta_{t+1})f)$, we have:
\begin{gather*}
\begin{aligned}
&(*)\xrightarrow{\text{P}}\sqrt{N}\tilde{\Omega}_{t}^{-1}\Sigma_{i=1}^{N} w_{t}^{(i)}\{a_{\theta_{t}}(\theta_{t+1})f(x_{0:t}^{(i)})-p_{\theta_{t}}(a_{\theta_{t}}(\theta_{t+1})f)\}\\
&\xrightarrow{\text{P}}\sqrt{N}\frac{p_{\theta_{t}}(y_{0:t})}{p_{\theta_{t+1}}(y_{0:t})}\Omega_{t}^{-1}\Sigma_{i=1}^{N} w_{t}^{(i)}\{a_{\theta_{t}}(\theta_{t+1})f(x_{0:t}^{(i)})-p_{\theta_{t}}(a_{\theta_{t}}(\theta_{t+1})f)\}\\
&\xrightarrow{\text{D}}N(0,(\frac{p_{\theta_{t}}(y_{0:t})}{p_{\theta_{t+1}}(y_{0:t})})^{2}{\sigma}^{2}(af)),
\end{aligned}
\end{gather*}
which proves the first statement. Here we used the fact that $f \in A_{t}$ implies $af\in A_{t}$ (similarly for $W_{t}$), which is straightforward by combining Assumption A with the construction of $\{A_{k},W_{k}\}_{k=0}^{t}$.

Then simply notice that $|a^{2}f|\leq(u_{2})^{2}|f|$ and thus 
\begin{gather*}
\begin{aligned}
&N\tilde{\Omega}_{t}^{-2}\Sigma_{i=1}^{N} (\tilde{w}_{t}^{(i)})^{2}f(x_{0:t}^{(i)})\\
\xrightarrow{\text{P}}&(\frac{p_{\theta_{t}}(y_{0:t})}{p_{\theta_{t+1}}(y_{0:t})})^{2}N\Omega_{t}^{-2}\Sigma_{i=1}^{N} (w_{t}^{(i)})^{2}a_{\theta_{t}}(\theta_{t+1})^{2}f(x_{0:t}^{(i)})\\
\xrightarrow{\text{P}}&(\frac{p_{\theta_{t}}(y_{0:t})}{p_{\theta_{t+1}}(y_{0:t})})^{2}{\gamma}(a^{2}f).    
\end{aligned} 
\end{gather*}
Finally, notice that: 
\begin{equation*}
\sqrt{N}\tilde{\Omega_{t}}^{-1}\max_{1\leq i\leq N}\tilde{w}_{t}^{(i)}\leq u_{2}\sqrt{N}\tilde{\Omega_{t}}^{-1}\max_{1\leq i\leq N}w_{t}^{(i)}\ \xrightarrow{\text{P}}u_{2}\frac{p_{\theta_{t}}(y_{0:t})}{p_{\theta_{t+1}}(y_{0:t})}\sqrt{N}\Omega_{t}^{-1}\max_{1\leq i\leq N}w_{t}^{(i)}\xrightarrow{\text{P}}0,
\end{equation*} 
which completes the proof.
\end{proof}

\begin{proof}[Proof of Theorem 1]
Recall that the semiGA-PIS algorithm proceeds as propagation, reweighting, retargeting and resampling. The result from \cite{chopin2004central} gives the influence of propagation, reweighting and resampling, while lemma 2 specifies the expression of variance $\bar{V}_{t}(f)$($\tilde{\sigma}^{2}(f)$in the setting of theorem 2) after retargeting, combining these results gives the four corresponding variances of the theorem. We now prove the expression for $V_{t}(f)$ by induction.

 For $t=0$ the statement obviously holds, suppose it holds for $t$, our calculation proceeds by following the inductive definitions. Notice that $f$(and also its domain) changes when the subscript time index increases, but we retain the notation $f$ to avoid overloading notations.
\begin{equation*}
\begin{aligned}
&\bar{V}_{t}(f)=(\frac{(p^{t}(y_{0:t}))^{2}}{p_{t+1}(y_{0:t})})^{2}V_{t}(a_{t}^{t+1}f)\\
&=\int \frac{(\pi_{t}^{t}(x_{0}))^{2}}{q(x_{0})}(\int \frac{\pi_{t+1}^{t}(x_{0:t})}{\pi_{t}^{t}(x_{0:t})}f(x_{0:t})\pi_{t}^{t}(x_{1:t}|x_{0})dx_{1:t}-\mathbb{E}_{\pi_{t+1}^{t}}(f(x_{0:t})))^{2}dx_{0}\\
&+\Sigma_{k=1}^{t}\int\frac{(\pi_{t}^{t}(x_{0:k}))^2}{\pi_{k}^{k-1}(x_{0:k})q(x_{k}|x_{0:k-1})}[\int \frac{\pi_{t+1}^{t}(x_{0:t})}{\pi_{t}^{t}(x_{0:k})}f(x_{0:t})dx_{k+1:t}-\int \pi_{t+1}^{t} f(x_{0:t})dx_{0:t}]^{2}dx_{0:k}.
\end{aligned}
\end{equation*}
Then
\begin{equation*}
\begin{aligned}
&\;\tilde{V}_{t+1}(f)=\bar{V}_{t}(\int q(x_{t+1}|x_{0:t})f(x_{0:t+1})dx_{t+1})+\text{Var}_{\pi_{t+1}^{t}}(f) +\mathbb{E}_{\pi_{t+1}^{t}}(\text{Var}_{q_{t+1}}(f))\\
&=\bar{V}_{t}(\int q(x_{t+1}|x_{0:t})f(x_{0:t+1})dx_{t+1})+\text{Var}_{\pi_{t+1}^{t}q_{t+1}}(f)\; (law\; of\; total\; variation)\\
&=\int \frac{(\pi_{t}^{t}(x_{0}))^{2}}{q(x_{0})}(\int \frac{\pi_{t+1}^{t}(x_{0:t})}{\pi_{t}^{t}(x_{0:t})}q(x_{t+1}|x_{0:t})f(x_{0:t+1})\pi_{t}^{t}(x_{1:t}|x_{0})dx_{1:t+1}-\int\pi_{t+1}^{t}q(x_{t+1}|x_{0:t}) f(x_{0:t+1})dx_{0:t+1})^{2}dx_{0}\\
&+\Sigma_{k=1}^{t}\int \frac{(\pi_{t}^{t}(x_{0:k}))^2}{\pi_{k}^{k-1}(x_{0:k-1})q(x_{k}|x_{0:k-1})}[\int \frac{\pi_{t+1}^{t}(x_{0:t})}{\pi_{t}^{t}(x_{0:k})}q(x_{t+1}|x_{0:t})f(x_{0:t+1})dx_{k+1:t+1}\\
&-\int \pi_{t+1}^{t}(x_{0:t})q(x_{t+1}|x_{0:t})f(x_{0:t+1})dx_{0:t+1}]^{2}dx_{0:k}\\
&+\int \pi_{t+1}^{t}(x_{0:t})q(x_{t+1}|x_{0:t})[f(x_{0:t+1})-\int \pi_{t+1}^{t}(x_{0:t})q(x_{t+1}|x_{0:t})f(x_{0:t+1})dx_{0:t+1}]^{2}.\\
\end{aligned}
\end{equation*}
Therefore
\begin{equation*}
\begin{aligned}
& V_{t+1}(f)=\tilde{V}_{t+1}(\frac{\pi_{t+1}^{t+1}(x_{0:t+1})}{\pi_{t+1}^{t}(x_{0:t})q(x_{t+1}|x_{0:t})}[f(x_{0:t+1})-\int \pi_{t+1}^{t+1}(x_{0:t+1})f(x_{0:t+1})dx_{0:t+1}])\\
& \stackrel{(**)}{=}\int \frac{(\pi_{t+1}^{t+1}(x_{0}))^{2}}{q(x_{0})}(\int \frac{\pi_{t+1}^{t+1}(x_{0:t+1})}{\pi_{t+1}^{t+1}(x_{0})}f(x_{0:t+1})dx_{1:t+1}-\int \pi_{t+1}^{t+1}f(x_{0:t+1})dx_{0:t+1})^{2}dx_{0}\\
&+\Sigma_{k=1}^{t}\int \frac{\pi_{t+1}^{t+1}(x_{0:k})}{\pi_{k}^{k-1}(x_{0:k-1})q(x_{k}|x_{0:k-1})}[\int \pi_{t+1}(x_{k+1:t+1}|x_{0:k})f(x_{0:t+1})dx_{k+1:t+1}\\
&-\int \pi_{t+1}^{t+1}(x_{0:t+1})f(x_{0:t+1})dx_{0:t+1}]^2dx_{0:k}\\
&+\int \frac{(\pi_{t+1}^{t+1}(x_{0:t+1}))^2}{\pi_{t+1}^{t}(x_{0:t})q(x_{t+1}|x_{0:t})}(f(x_{0:t+1})-\int \pi_{t+1}^{t+1}(x_{0:t+1})f(x_{0:t+1})dx_{0:t+1})^{2}dx_{0:t+1}\\
&=\int \frac{(\pi_{t+1}^{t+1}(x_{0}))^{2}}{q(x_{0})}(\int f(x_{0:t+1})\pi_{t+1}^{t+1}(x_{1:t+1}|x_{0})dx_{1:t+1}-\mathbb{E}_{\pi_{t+1}^{t+1}}(f(x_{0:t+1})))^{2}dx_{0}\\
&+\Sigma_{k=1}^{t+1}\int \frac{\pi_{t+1}^{t+1}(x_{0:k})}{\pi_{k}^{k-1}(x_{0:k-1})q(x_{k}|x_{0:k-1})}[\int \pi_{t+1}(x_{k+1:t+1}|x_{0:k})f(x_{0:t+1})dx_{k+1:t+1}\\
&-\int \pi_{t+1}^{t+1}(x_{0:t+1})f(x_{0:t+1})dx_{0:t+1}]^2dx_{0:k},
\end{aligned}
\end{equation*}
where$(**)$ includes simplification$\int \pi_{t+1}^{t+1}[f(x_{0:t+1})-\int \pi_{t+1}^{t+1}f(x_{0:t+1}dx_{0:t+1})]dx_{0:t+1}=0.$ 
\end{proof}

\section{MORE DETAILS AND RESULTS ON THE NUMERICAL EXPERIMENTS}
\hypertarget{2}
All numerical experiments were implemented in Python 3.0. For the AR(1) with noise model, we use the optimal proposal for SMC. Standard bootstrap methods are applied for all algorithms in the rest of the models, and for the semiGA-PIS algorithm, we choose $K=1$ for $ESS_{k}$.  The compute worker is CPU: Processor 12th Gen Intel(R) Core(TM) i7-12700 2.10 GHz,
with installed RAM 32GB, under the 64-bit operating system, x64-based processor. We use the bootstrap SMC, i.e. vanilla SMC in Algorithm 1 with the proposal $q_{\theta}(x_{t+1}|x_{0:t})=f_{\theta}(x_{t+1}|x_{t})$. The performance of the algorithms is measured by the Root Mean Squared Error(RMSE):
\begin{equation}
    \label{rmse}
    RMSE(\theta;\theta_{t}^{1:S})\triangleq \sqrt{S^{-1}\Sigma_{s=1}^{S}(\theta_{t}^{s}-\theta)^{2}},
\end{equation}
where superscript $s$ indicates the $s^{th}$ among the total of $S$ experiments and subscript $t$ is the index of the iteration, $\theta$ refers to the value that RMSE is calculated for, usually set to be $\theta_{true}$ or $\hat{\theta}_{MLE}$.

\hypertarget{nan}Python sometimes returns $nan$ value due to the failure of the algorithm, which may come from division by very small values close to zero, or improper command asking the density of $N(a,b)$ with $b<0$, etc. We point this out because when the $nan$ value is returned we could not use the formula for RMSE anymore. Specifically, for the illustration of robustness in section 5.1(where the parameter of interest$\theta$ corresponds to $\phi$), we address this by only calculating RMSE when $1.3>\theta_{t}^{s}>0.6$, and otherwise adding a mild penalty of $0.35^{2}$ on the term.

\hypertarget{appendixD}
The following are discussions for the choices of learning rates in all models based on the guidelines from \citet{spall1998implement}. Specifically $\gamma_{t}$ is in the form of $\frac{c_{1}}{(A+t)^{\alpha}}$. The finite difference method for Naive-SGA uses SPSA\citep{spall1998implement} where the perturbation vector is  $\tau_{t}\Delta_{t}$ where$\tau_{t}\triangleq \frac{c_{2}}{(A+t)^{\beta}}$ and $\Delta_{t}$ is a Rademacher random vector(i.e. each component taking 1 or -1 with probability 0.5) having the same dimensions as the parameter of interest $\theta$. We choose the pair $(\alpha,\beta)=(1,1/6)$, which is an optimal choice based on a large number of experiments\citep{spall1998implement}. For model\eqref{ar1} we choose $(c_{1},c_{2},A)=(0.0001,0.05,100)$. For model\eqref{sv} we choose $(c_{1},c_{2},A)=(0.00005,0.05,100)$. For the PAR(1) model we choose $(c_{1},c_{2},A)=(0.2,0.02,2000)$, we choose large $A$ here because the data size $T=168$ is small compared to the previous experiments with $T=10000$ and we want to reduce the effect of wrong updates for the online algorithm based on early iterations with possibly outlier observations. The joint choice of $c_{1}, A$ in all models are based on the magnitude of the gradient and the scale of the parameter. Particularly, we would like to address the fact that the learning rates are reasonable and fair in the sense that their scales are sufficient for multiple algorithms to converge, given that the gradient estimations of the algorithms are correct. More details are contained in the code.

\subsection{Tuning the ESS thresholds $r$ and $r_{1}$}

Essentially this is done by finding reasonable values of ESS in test runs of the algorithms which use smaller data size. Specifically, for the offline estimation, multiple parameter values are selected randomly in a region of reasonable parameter values. By moving each of them with one Newton-Raphson step size which can easily be estimated with the Fisher and Louise identities, a set of ESS values are calculated. The average of these values represents the ESS when the step size is optimal and is chosen as $r$.

For online estimation, in the test run $\theta_t$ is moved by a fixed value of the distance $d$, instead of the scaled conditional gradient in Algorithm 3, at every $t$ and the value of ESS is recorded. Also in the test run particles are renewed at a relatively low frequency. By doing this we observe the range of ESS which includes the values when the move ranges from too small to too far, and the medium of the recorded ESS is used as $r_1$.  Several values of $d$ are tested by taking the physical meaning of the parameter into consideration, where values of $d$ only giving small or large values of ESS are excluded.  

We have tried different values of $r$ and $r_1$ for the SV model in Section 4.2 where $r_1$ is obtained by using different $d$ in the above tuning procedure, and the results are similar which is given below. 


\begin{table}[h]
  \caption{Experimenting different $r, r_{1}$ in section 4.2}
  \label{ex5}
  \centering
  \begin{tabular}{llll}
    \toprule
    Algorithms     & $\phi$  &$\sigma_{x}$ &$\sigma_{y}$ \\
    \midrule
    P-offline $O(N^{2})$ &2.403  &2.031 &0.220\\
    P-online $O(N^{2})$ &1.000 &1.000 &1.000\\
    Naive SGA &4.552 &2.512 &2.023  \\
    Fisher SGA     &3.925 &1.673 &0.896 \\
    AdaptGA-PIS($r=0.4$) &0.151 &0.802 &0.048\\
    \bottomrule
    \end{tabular}
      \begin{tabular}{llll}
    \toprule
        Algorithms     & $\phi$  &$\sigma_{x}$ &$\sigma_{y}$ \\
    \midrule
    AdaptGA-PIS($r=0.334$) &0.343 &0.892 &0.059\\
    AdaptGA-PIS($r=0.449$) &0.139 &0.727 &0.045\\
    SemiGA-PIS($r_1=0.6$) &0.235 &0.265 &0.059  \\
    SemiGA-PIS($r_1=0.487$) &0.235 &0.237 &0.045 \\
    SemiGA-PIS($r_1=0.692$) &0.241 &0.285 &0.074\\

    \bottomrule
  \end{tabular}
\end{table}
 The result shows that the performance of our algorithms is satisfying with $r$ and $r_{1}$ tuned by the above procedure, and is reasonably insensitive to the choice of $r$ and $r_{1}$.


\subsection{The Naive and Fisher SGA Algorithms}
We present the structure of the two SGA algorithms.
\begin{algorithm}[H]
\caption{Steepest gradient ascent(SGA) using SMC}
  \begin{algorithmic}
   \State
 Assume that $y_{0:T}$,  the number of total iterations $I$, step number $K$ and properly chose $\{\gamma_{n}\}_{n=0}^{I}$ are given.\\
\noindent\textbf{Initialisation}
Specify the initial parameter $\theta_{0}$.\\
\noindent \textbf{For} $n = 0$ to $I$:
\begin{adjustwidth}{1cm}{0cm}
1. Run the SMC algorithm to obtain $\{x_{0:T}^{(i)},w_{T}^{(i)}\}_{i=1}^{N}$ follow the density $p_{\theta_{n}}(x_{0:T}|y_{0:T})$.\\
2. Use finite difference or Fisher's identity to estimate $\nabla \log p_\theta\left(y_{0: t}\right)|_{\theta=\theta_{n}}$, and proceed gradient ascend $\theta_{n+1}=\theta_n+\gamma_{n} \nabla_{\theta} \log p_\theta\left(y_{0: T}\right)|_{\theta=\theta_{n}}$
\end{adjustwidth}
\textbf{End For}
   \end{algorithmic}
\end{algorithm}
The algorithm has a similar structure as our proposed offline algorithm, but simulating new particles at every iteration.

\subsection{Observed Datasets in Section 4.1 and 4.2}
\begin{figure}[h]
  \centering
 \includegraphics[width=0.3\textwidth]{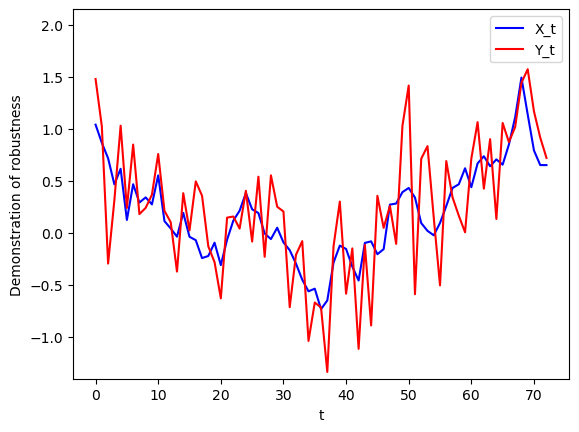}
\includegraphics[width=0.3\textwidth]{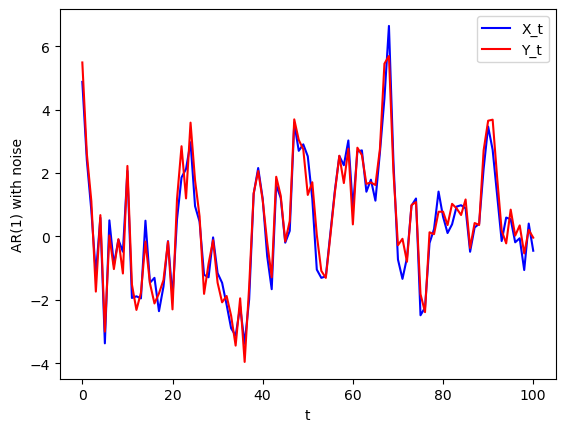}
   \includegraphics[width=0.3\textwidth]{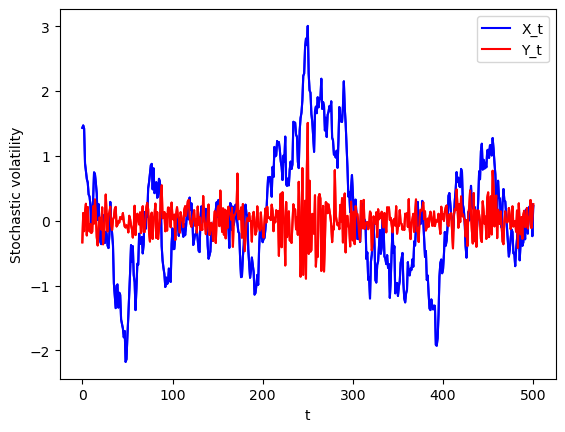}
  \caption{The observed dataset simulated from model \eqref{ar1_changed} (left), model \eqref{ar1}(middle) and model \eqref{sv}(right)}
\label{obs_data}
\end{figure}

Here in Figure \ref{obs_data} we present the plots for the observed data used in Section 5. The real dataset in Section 5.3 is not presented here.

\subsection{More Results for Section 4.1} 

\begin{figure}[h]
  \centering

\includegraphics[width=0.4\textwidth]{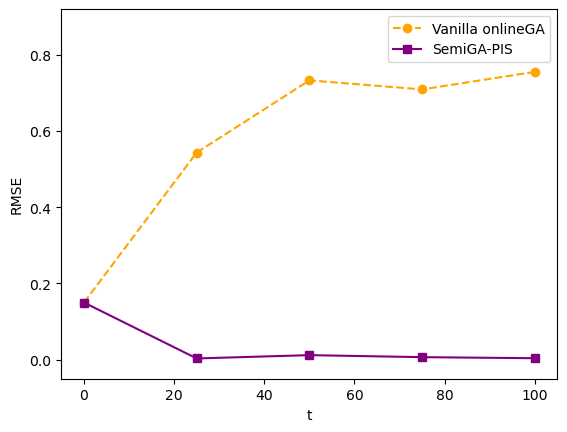}
\includegraphics[width=0.4\textwidth]{onlinefailure1.png}
  \caption{Comparison of Algorithm \ref{onlineMLE}(vanilla onlineGA) and Algorithm \ref{onlineAIS}(semiGA-PIS) model \eqref{ar1_changed} with $\phi_\star=0.95$ and the initial parameter $\phi_{0}=0.8$. $\sigma_x=0.5$ and $\sigma_y=0.5$ are known. The RMSE plot(left) is obtained over $20$ replications, and the trajectory plot(right) is from one of the replications. }
\label{robust_appen}
\end{figure}

\begin{figure}[h]
  \centering
\includegraphics[width=0.4\textwidth]{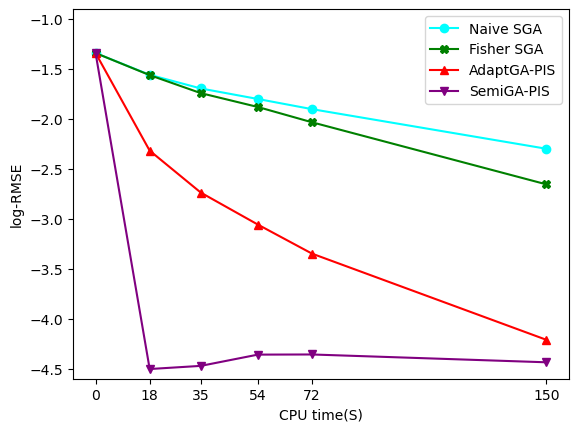}
\includegraphics[width=0.4\textwidth]{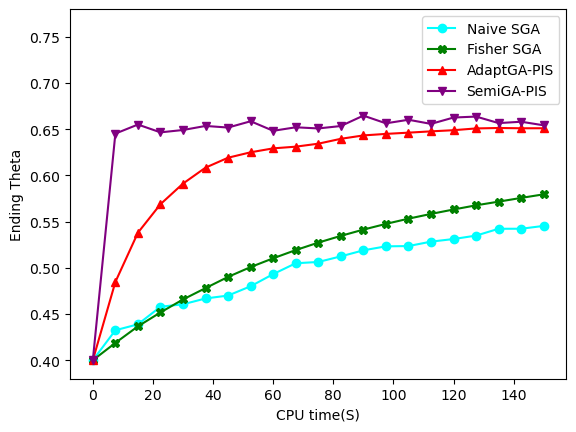} \\	
  \caption{Log-RMSE plot obtained with $20$ replications(left) and the trajectory plot from one replication (right) under model \eqref{ar1_changed}. The initial value $\phi_{0}=0.4$. The true parameter $\phi^{\star}=0.7$, and $\sigma_x=0.5$ and $\sigma_y=0.5$ are known.}
\label{ex1_2_appen}
\end{figure}

For model \eqref{ar1_changed}, the RMSE plot comparing Algorithm \ref{onlineMLE} and Algorithm \ref{onlineAIS} and the one comparing other algorithms are reported in Figure \ref{robust_appen} and \ref{ex1_2_appen}, respectively. These results show that the semi-online algorithm significantly outperforms the ad-hoc  online algorithm and the offline algorithms when part of the observed sequence is highly informative about the unknown parameter.

\subsection{More Results for Section 5.2}

\begin{figure}[h]
  \centering
 \includegraphics[width=0.3\textwidth]{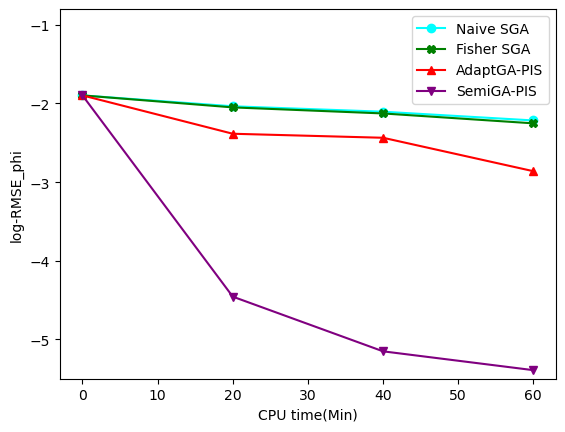}
\includegraphics[width=0.3\textwidth]{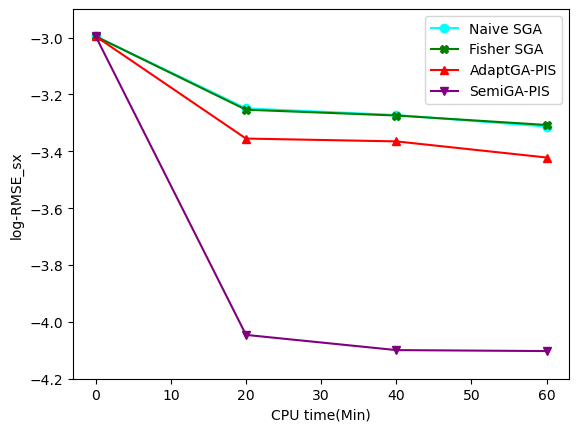}
   \includegraphics[width=0.3\textwidth]{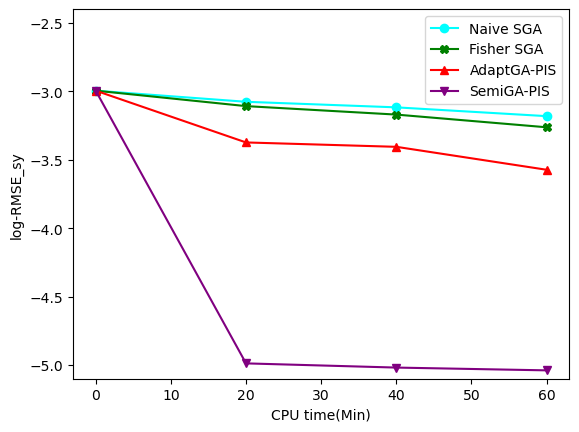}
  \caption{ Log-RMSE plots from 20 replications for $\phi$(left), $\sigma_{x}$(middle) and $\sigma_y$(right) for all algorithms under the SV model. The true parameter $\theta^{\star}=(0.95,0.25,0.2)$ and $T=500$. For all algorithms $N=2000$, the initial parameters $\theta_{0}=(0.8,0.2,0.15)$,  $r=0.6$ for Algorithm \ref{offline_algo} and $r_{1}=0.8$ for Algorithm \ref{onlineAIS}.}
\label{ex2_appen} 
\end{figure}

Consider the following stochastic volatility model where a degenerating trend term is added to the mean volatility:
\begin{equation*}
\begin{aligned}   
    & X_{t+1}\sim \phi X_{t}+\sigma_x\eta_{t},\; t=0,\dots, T-1,\\
    & Y_{t}\sim \sigma_{y}e^{(8\phi^t+X_{t})/2}\xi_{t},\; t=0,\dots, T,\\
\end{aligned}
\end{equation*}
where $X_0$ has mean $0$ and $\eta_{t}$ and $\xi_{t}$ follow $N(0,1)$. The parameter to estimate is $\theta=(\phi,\sigma_{x},\sigma_{y})$. The MLE of $\theta$ is obtained by running all algorithm with large values of $N$ and iteration numbers until sure convergence and the converged value is agreed by the majority of algorithms.
The RMSE plots for all three parameters are shown in Figure \ref{ex2_appen}. Again the semi-online algorithm significantly outperforms all other algorithms for a similar reason as before. The naive SGA performs the worst, with the Fisher SGA only slightly improving over it.

\end{document}